\documentclass[preprint2]{aastex}
\usepackage{natbib,float}

\usepackage{color}

\shorttitle{Emission Lines in Mrk 231}
\shortauthors{Leighly et al.}

\begin{document}


\title{ The Binary Black Hole Model for Mrk 231 Bites the Dust}


\author{Karen M. Leighly\footnote{Visiting Astronomer at the Infrared
    Telescope Facility, which is operated by the University of Hawaii
    under Cooperative Agreement no. NNX-08AE38A with the National
    Aeronautics and Space Administration, Science Mission
    Directorate, Planetary Astronomy Program.}}
\affil{Homer L.\ Dodge Department of Physics and Astronomy, The
  University of Oklahoma, 440 W.\ Brooks St., Norman, OK 73019}

\author{Donald M. Terndrup}
\affil{Department of Astronomy, The Ohio State University, 140
  W.\ 18th Ave., Columbus, OH 43210} 

\author{Sarah C. Gallagher}
\affil{Department of Physics \& Astronomy and Centre for Planetary and
  Space Exploration, The University of Western
  Ontario, London, ON, N6A 3K7, Canada}

\author{Adrian B.\ Lucy}
\affil{Department of Astronomy, Columbia University, 550 W.\ 120th
  Street, New York, NY 10027}




\begin{abstract}

Mrk~231 is a nearby quasar with an unusually red near-UV-to-optical
continuum, generally explained as heavy reddening by dust
\citep[e.g.,][]{leighly14}. \citet{yan15} proposed that Mrk~231 is a
milli-parsec black-hole binary with little intrinsic reddening. { We
show that if the observed FUV continuum is intrinsic, as assumed by
\citet{yan15}, it fails by a factor of about 100 in powering the
observed strength of the near-infrared emission lines, and the thermal
near and mid-infrared continuum.  In contrast, the line and continuum
strengths are typical for a reddened AGN spectral energy distribution.}
We find that the \ion{He}{1}*/P$\beta$ ratio is sensitive to the
spectral energy distribution for a one-zone model.  If this
sensitivity is maintained in general broad-line region models, then
this ratio may prove a useful diagnostic for heavily reddened
quasars. Analysis of archival {\it HST} STIS and FOC data revealed
evidence that the far-UV continuum emission is resolved on size
scales of $\sim 40$ parsecs.  The lack of broad absorption lines in
the far-UV continuum might be explained if it were not coincident with
the central engine.  One possibility is that it is the central engine
continuum reflected from the receding wind on the far side of the
quasar. 

\end{abstract}


\section{Introduction\label{intro}}

The confirmed existence of a milli-parsec-separation supermassive
black hole (SMBH) binary would be an important discovery.  SMBH
binaries may be present in nature as a consequence of
hierarchical mergers of dark matter halos, so the incidence of binary 
AGN provides a potentially important test of galaxy assembly models.
They may be a strong source of gravitational wave emission,  and
are therefore potentially important probes of general relativity. 

Mrk~231 is a well-known, nearby ($z=0.0421$) ultra-luminous infrared
galaxy that has a Seyfert 1 optical spectrum \citep{sanders88}. The
infrared emission is thought to be a combination of AGN and starburst
activity \citep[e.g.,][and references therein]{farrah03}.  Recently,
attention has been again drawn to this galaxy after the discovery of a
powerful, wide-angle, kiloparsec-scale molecular outflow
\citep{rupke11}.    

Mrk~231 has an unusual spectral energy
distribution (SED).  While the optical through infrared SED appears
typical of quasars, the spectrum is strongly cut off through the near
UV, and the continuum is very weak toward shorter wavelengths.  The
near-UV-through-infrared SED was studied by \citet{leighly14}.  We
found that the unusual shape was consistent with circumstellar
reddening, which is distinguished by a large covering fraction, and
large optical depths, approaching one, that produce increased
extinction in the blue and UV, with red light scattered back into the
line of sight as a secondary effect.  This type of reddening has been
observed in Type 1a supernovae \citep{wang05,
  goobar08}{, and in Gamma-ray Burst afterglows
  \citep{fynbo2014}. Some of the theory was developed to treat the
  general case of the transfer of radiation in galaxies \citep{witt92}.
  In a systematic study of quasar reddening performed by 
  \citet{krawczyk15}, while most SEDs were best fit by a SMC
  reddening curve, a few were better fit by the circumstellar
  one. This type of   reddening is natural when the geometry is
  predominately spherical, rather than a screen as usually assumed.}
\citet{veilleux13} also interpreted the 
unusual near-UV-through-optical spectral energy distribution in terms
of reddening, although their proposed reddening mechanism was somewhat 
different.  

\begin{figure*}[!t]
\epsscale{1.0}
\begin{center}
\includegraphics[width=5.5truein]{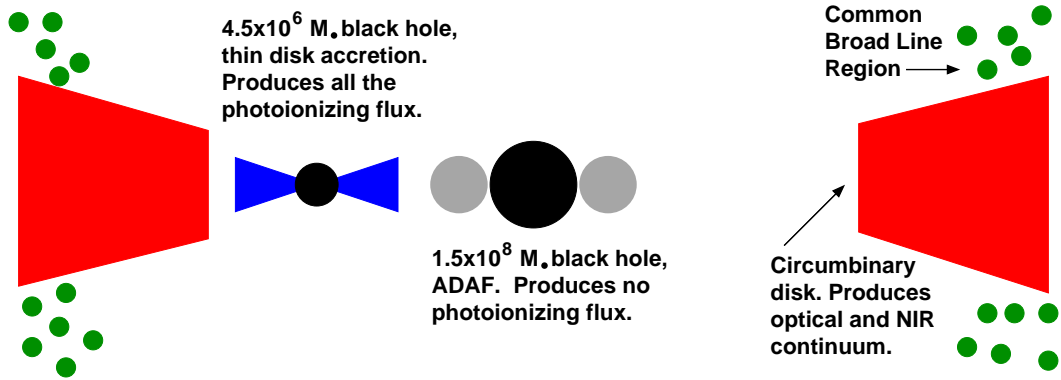}
\caption{A schematic diagram of the \citet{yan15} binary black hole
  model for Mrk~231; see also \citet[][Figure 1]{yan15} for a face-on
  view.  The smaller $4.5\times 10^6 \rm \, M_\odot$ black hole
  accretes as a thin disk and radiates efficiently, thereby emitting
  all the photoionizing flux that powers the broad-line region
  emission, while the larger $1.5 \times 10^8 \rm \, M_\odot$ black
  hole accretes at a low rate, as an ADAF, and thus radiates
  inefficiently, producing no significant photoionizing
  flux. The BLR is predicted to be about 1.5 times the radius of the
  inner edge of the circumbinary disk, and is required emit at more
  than   100 times the flux of a typical BLR to produce the observed
  near-infrared emission lines.  \label{fig1}}   
\end{center}
\end{figure*}

\citet{yan15} proposed an alternative explanation of the unusual
SED. They suggest that Mrk 231 hosts a milli-parsec binary black hole
system, with nearly negligible reddening.  The smaller-mass black hole
($4.5 \times 10^6 \rm \, M_\odot$) accretes as a thin disk and
dominates the weak UV  emission.  The larger-mass black hole
($1.5\times 10^8 \rm \, M_\odot$) has a low accretion rate and
radiates inefficiently as an Advection Dominated Accretion Flow
(ADAF).  These two black holes are surrounded by a circumbinary disk,
which dominates the optical and IR, and the steep rolloff observed
toward the UV is the inner edge of the circumbinary disk.  Moreover, they suggest that a steep rolloff
from the optical toward the UV is a characteristic signature of binary
black holes, and thus finding objects with similar spectra provides a
method to discover these objects.  A not-to-scale schematic diagram of
the model is shown in Fig.\ref{fig1}; see also \citet{yan15} Fig.\ 1.

In this paper, we critique the binary black hole model for Mrk~231
presented by \citet{yan15};  in particular, we examine whether the
binary black hole model can produce the observed emission lines.  
In \S\ref{model_review}, we review the features and assumptions of
the \citet{yan15} model relevant to the production of the emission
lines.  \S\ref{data}  motivates our use of near-infrared emission
lines, combined with \ion{C}{4}, as diagnostics, and describes the
data that we compiled.  \S\ref{sed} lays out the several spectral 
energy distributions used for simulations.    \S\ref{cloudy} reports
the results obtained using a set of one-zone {\it Cloudy}
models.  \S\ref{fuv} describes the analysis of archival {\it HST} STIS
and FOC data indicating evidence for extended emission in the far UV.
\S\ref{discussion} summarizes our results: we find that 
the binary black hole model cannot produce the observed emission
lines.  We also show that the binary black hole model cannot power the
observed mid-infrared continuum (torus) emission.  We briefly discuss
the potential utility of the infrared emission lines for constraining
the intrinsic spectral energy distribution in obscured quasars, as
well as a possible reflection origin for the far UV continuum.

\section{The \citet{yan15} Model\label{model_review}}

A critical feature of the  \citet{yan15} model is that only
the emission from accretion onto the small mass black hole contributes
significantly to the photoionizing continuum.  They identify the
continuum of the small-mass black hole with the weak far-UV continuum
emission 
\citep{veilleux13}, while the near-UV through infrared emission
originates in the circumbinary disk.   Thus, their model requires that
the photoionizing flux of a small-mass black hole power the line
emission of a quasar with a combined mass of $\sim 2$ orders of
magnitude larger.  As we will show in \S\ref{mrk231}, if this were
true, then the observed near-infrared lines would be required to 
exhibit equivalent widths with respect to the photoionizing continuum 
(rather than the observed continuum) about 100 times larger than
normal.   

\citet{yan15} make assumptions about the spectral energy distribution
of the small mass black hole that result in a relatively high fraction
of its emission emerging in the far UV.  Inferred to be accreting at a
significant fraction of Eddington, the small black hole's accretion
disk would be expected to be bright, and have a high inner-edge
temperature.   They assume that the 
optical-UV continuum is very blue, having the slope of the
sum-of-blackbodies accretion disk, i.e., $F_\nu \propto \nu^{1/3}$.
In contrast, quasars are generally observed to have a redder
optical-UV continuum, closer to $F_\nu \propto \nu^{-0.5}$; this is
one of the unexplained mysteries of accretion disks, along with the
lack of polarization and clear evidence for Lyman edges  \citep{kb99}.
Finally, they assume that the X-ray flux is negligible.  This is 
inconsistent with observations, as we show in \S\ref{sed}. These
factors combine to make their assumed accretion disk emit a maximally
high fraction of its continuum in photoionizing photons.  We explore
the effect of their optimized spectral energy distribution the
predicted line emission in \S\ref{cloudy}, along with the predicted
emission using more typical SEDs, and show that none of them can
produce the observed flux and ratios of the near-infrared helium and
hydrogen broad lines. 

\citet{yan15} address the question of line emission in their \S
6.  They make a simple nebular approximation and compute the expected
number of H$\alpha$ photons resulting from recombination of hydrogen
in a gas illuminated by their assumed spectral energy distribution.
They predict that sufficient H$\alpha$ emission would be produced if
the global covering fraction is $\Omega \sim 0.5$.   However, this
analysis is insufficient, because hydrogen
lines are produced in the partially-ionized zone in quasars, and the
line ratios and fluxes are observed to be different from
those predicted using the simple nebular approximation
\citep[e.g.,][]{of06,   kk81, dn79}.  Also, quasars produce other emission
lines besides the Balmer lines, and it is not obvious that those would
be produced with sufficient strength to match observations.  We
perform a more realistic and complete line analysis in \S\ref{cloudy}.     

\section{Data\label{data}}

\subsection{Emission Lines Considered\label{emlines}}

Our goal in this paper was to determine whether the line emission
observed in Mrk 231 is consistent with the binary hypothesis put forth
by \citet{yan15}, not to provide a full model of the emission lines.
Therefore we considered just a few lines that provide sufficient
diagnostic power. 

The near infrared spectrum is 
relatively free of the effects of reddening, regardless of the
interpretation of the continuum.  Our spectrum was obtained using the
SpeX instrument on the IRTF, and the details of the observation are 
found in \citet{leighly14}.  We use \ion{He}{1}*$\lambda 10830$, a
line that arises from recombination of once-ionized helium.   The
energy required to create He$^+$ is 24~eV, and therefore \ion{He}{1}*
probes the \ion{H}{2} part of the broad-line region emission.  Being a
recombination line, it is  principally a diagnostic of the flux of the
ionizing continuum on the broad-line-region gas.  For example, for a
semi-infinite slab, the flux of this line is monotonic with the
helium-continuum photoionizing  flux.

We also used Paschen $\beta$ at 12818\AA\/ and Paschen $\alpha$ at
18751\AA\/.   Paschen $\beta$ occurs in close proximity to the 
\ion{He}{1} line in the spectrum, so reddening does not affect their
line ratios significantly.  The ratio of Paschen $\alpha$ to Paschen
$\beta$ can be influenced by reddening, but much less than, for
example, the Balmer lines, since reddening curves flatten
toward the near infrared.   

The Paschen lines are recombination lines of hydrogen, and their
fluxes and ratios are influenced strongly by the physical
conditions of the line-emitting gas.  Although these
lines are produced throughout the ionized gas slab, a significant
amount  is produced beyond the hydrogen ionization front, in the
partially-ionized zone.  Our simulations (\S \ref{cloudy})  show that
a significant column of partially-ionized gas is required for these 
lines to be observable against the bright quasar
continuum.  In the partially-ionized zone, the opacity to Lyman 
lines can be very large, and hydrogen line ratios are dramatically
different from those predicted in the simple nebular approximation
\citep[e.g.,][]{of06, kk81}.   For example, absorbed and thermalized
Ly$\alpha$ can create a significant population of hydrogen in $n=2$,
which can then suffer photoionization by photons with wavelengths
shorter than 3646\AA\/.  X-ray photoionization is also important.   In
addition, while hydrogen in $n=1$ cannot be 
collisionally excited because the difference in energy between $n=1$
and $n=2$ is too large, hydrogen in $n=2$ can experience collisional
excitation, and this process will contribute to the Balmer and Paschen 
lines.  The presence of a population of hydrogen in $n=2$ means that
Balmer lines will also experience significant optical depths, reducing
the radiative de-excitation and cooling.  Additional processes such as
charge exchange and collisional de-excitation may also
contribute. Turbulence or differential velocities will change the line 
optical depth, further altering the line ratios
\citep[e.g.,][]{bottorff00}. This means that the recombination line
fluxes and  ratios are best estimated using a photoionization code
such as {\it   Cloudy}, which accounts for all of these processes. 

We use \ion{C}{4}$\lambda \lambda 1548, 1551$ to probe the
UV where reddening is important.  \ion{C}{4} is a
collisionally-excited line that is produced in the \ion{H}{2} portion
of the broad-line region.  It is generally one of the strongest lines
present in quasar spectra, a consequence of the relatively high
abundance of carbon, and its easy excitability, as C$^{+3}$ has only
one valence electron.    

In this paper, we use as our diagnostics the strength of the
\ion{He}{1}* line (either the equivalent width of the line, or the
predicted flux, depending on the situation; see below), and the line
flux ratios \ion{He}{1}*/P$\beta$, P$\alpha$/P$\beta$, and
\ion{He}{1}*/\ion{C}{4}.  Thus, we are leaving out a great deal of
physics known to be relevant to the broad line region.  For example,
we do not take into account the fact that the BLR is not likely to be
characterized by a single set of physical parameters (ionization
parameter, density, column density) but rather to be a superposition
of line emission from gas characterized by a range of parameters.
This latter case is addressed by the Locally Optimally Emitting Clouds
(LOC) models \citep{baldwin95}, and LOC models for some of the lines
considered in this paper have been investigated by \citet{ruff12}.  In
this paper we consider a one-zone photoionization model, principally
because it is sufficient to prove our point, as the effect that we see
is very apparent, even without photoionization analysis, as described
in \S\ref{mrk231}.    

It is also known that various emission lines can have different shapes 
that reflect an origin in kinematically different gases.  For example,
in some relatively rare cases, the high-ionization lines such as
\ion{C}{4} can be broad and blueshifted, while intermediate-
and low-ionization lines are narrow and symmetric about the rest
wavelength \citep[e.g.,][]{leighly04}.  We do
not address this in our analysis either,  again because the result we
find is not subtle, and because the observed line profiles don't 
warrant this consideration.

Finally, emission lines are known to respond to the shape of the
spectral energy distribution  (SED). If the SED is hard, i.e., the
X-ray band is strong relative to the UV, then high-ionization lines
are observed to be strong \citep[e.g.,][]{casebeer06}.  If the SED is
soft, then the high-ionization lines are observed to be weak
\citep[e.g.,][]{leighly07}.  We address this effect by considering
several spectral energy distributions that to first approximation span
the range of shapes observed in AGN and quasars.

\subsection{Mrk 231\label{mrk231}}

The \ion{He}{1}*, P$\beta$, and P$\alpha$ lines were modeled in
\citet{leighly14}, and we take those measurements from that analysis.    

{\it HST} has observed the  \ion{C}{4} region twice, and both
observations are available in the archive.  The first
observation was taken in 1996 with FOS and is shown in
\citet{gallagher02}.  The second one was taken in 2014 using COS.
They are consistent with one another, although the COS spectrum has
much better signal-to-noise ratio.  We measure the flux in the
\ion{C}{4} line in the COS spectrum by fitting it with two Gaussians. 

Analysis of the near-infrared line equivalent widths shows us, in a
simple qualitative way, why the \citet{yan15} model is untenable.  It
is known that the properties of the broad line emission among AGN and
quasars (e.g., the lines observed, their equivalent widths, and their
ratios) are roughly the same over a factor of $\sim 10,000$ in
inferred black hole masses and luminosities.  Some variation does
occur, for example, the Baldwin effect \citep{baldwin95}, but this
accounts for a 
variation of less than one order of magnitude in equivalent width over
four orders of magnitude in luminosity
\citep[e.g.,][Fig. 7]{dietrich02}.  The constancy of the line
equivalent widths means that photoionizing flux scales with the
continuum under the emission lines. Thus, luminous quasars have
luminous BLR emission, and less luminous AGN have less luminous BLR
emission. So it is not reasonable to expect that a small black hole
will be able to power the line emission of a luminous quasar.     

\begin{figure*}[!t]
\begin{center}
\epsscale{1.0}
\includegraphics[width=4.5truein]{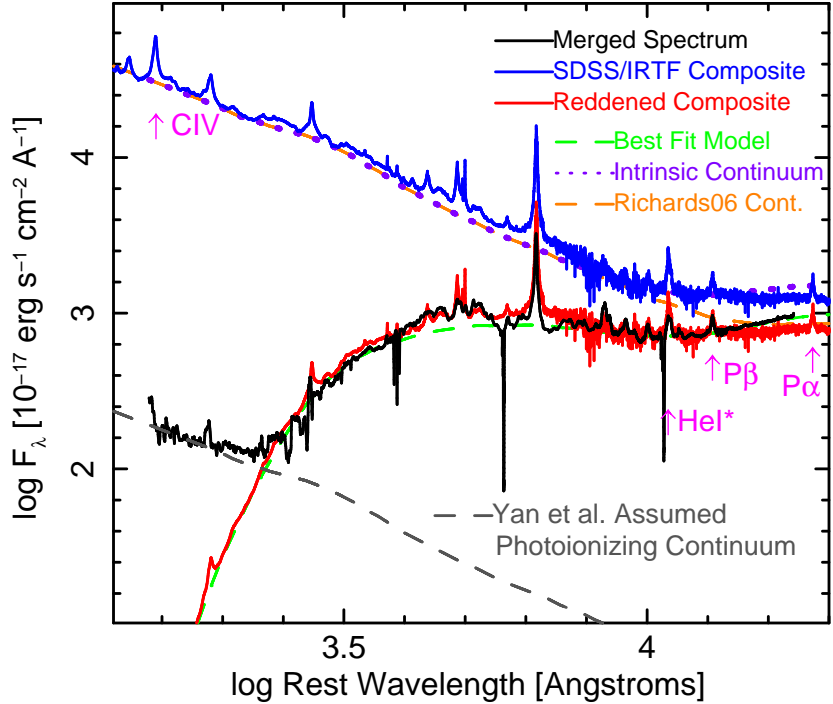}
\caption{Adapted from \citet{leighly14} Figure 5. We
show the observed broad-band spectrum from Mrk~231 (black)
overlaying the best-fitting continuum model (dashed green) developed
using the \citet{richards06} continuum (dashed orange) plus a
blackbody component from hot dust, and subject to the
circumstellar reddening;  see \citet{leighly14} for details.  Overlaid
is a composite spectrum created merging the SDSS quasar composite 
\citep{vandenberk01}   and the IRTF quasar composite
\citep{glikman06} in solid blue, and subject to the same reddening, in
solid red; see Section \ref{mrk231} for details.  This yields a good
fit over most of the bandpass, except for the continuum shortward of
$\sim 2200$\AA\/; this is a separate component that may arise from
scattering (\S \ref{spec}).   We also overlay the \citet{richards06}
continuum scaled to match the continuum flux at 2000\AA\/ (dashed
gray).  In the \citet{yan15} model, this taken to be the emission from
the accretion disk of the small-mass black hole. The wavelengths of
the four diagnostic lines used in this paper are  
marked.\label{fig2}}
\end{center}
\end{figure*}

This idea is investigated  qualitatively in Fig. \ref{fig2}, which
shows a variation of \citet{leighly14} Figure 5 that includes analysis
of the emission lines.  We first overlaid the intrinsic continuum,
inferred in \citet{leighly14}, with an optical-IR composite spectrum
created by joining the SDSS composite \citep{vandenberk01} to the
SDSS/IRTF composite  \citep{glikman06} at around 4000\AA\/.  The
merged composite provides infrared coverage, and avoids the well-known
long-wavelength flattening present in the SDSS quasar composite due to
host galaxy  contributions of lower-luminosity quasars that dominate
the SDSS composite on the red end of the visible band.  The match is
good except at very long wavelengths where an additional hot blackbody 
component is present \citep{leighly14}.     

Next, we applied the \citet{goobar08} reddening curve, using the
parameters that we inferred in \citet{leighly14}, to the quasar
composite spectrum.  Overlaid is the observed Mrk~231 spectrum.  The
overall agreement is very good through the optical and near UV where
the line emission is typical of quasars.  Mrk~231 has
somewhat stronger \ion{Fe}{2} emission, and somewhat weaker Balmer
emission, but the difference is within the range observed among AGN
and quasars. The reddened composite spectrum also provides a
reasonable match to the near-UV continuum, against which the
\ion{Mg}{2} and \ion{Fe}{2} broad  absorption lines expected in this
FeLoBAL can be seen.  We discuss the far-UV in
\S\ref{fuv}. 

The \citet{yan15} model posits that the continuum from the
normally-accreting small black hole is swamped by the circumbinary
disk emission through the optical and infrared, but it becomes visible
in the UV, shortward of the rolloff caused by the inner edge of the 
circumbinary disk.  The gray dashed line in Fig. \ref{fig2} shows the
\citet{richards06} continuum scaled to match the near-UV spectrum;
this represents the putative continuum of the small black hole.  In
the region of the infrared lines of interest, around 1 micron, this
continuum is a factor of $\sim 100$ below the observed Mrk 231
continuum.  {\it  Thus, if the small black hole is producing all of
  the   photoionizing photons in the system, then the infrared line
  equivalent widths would have to be a factor of $\sim 100$ stronger 
  than typical with respect to the small black hole continuum in order
  to show the observed equivalent widths with respect to the observed
  continuum.}  This does not seem plausible, as such huge equivalent
widths have never been seen. Indeed, as we will show in \S
\ref{yan_cloudy}, the smaller black hole continuum lacks sufficient
power to produce the observed infrared line emission.  

We used the inferred luminosities to estimate the radius of
the broad line region.  Interpolating the flux density at 5100\AA\/
from the \citet{leighly14} intrinsic continuum and the \citet{yan15}
photoionizing continuum (orange and gray lines in Fig.~\ref{fig2}), and using
the ``clean'' parameters for the radius/luminosity relationship from
\citet{bentz13}, the H$\beta$ emitting BLR is estimated to
be located 99 and 6.5 light days from the central engine for the 
reddened and the binary black hole interpretations, respectively.  The
first value seems typical for a quasar, while the second value is
small among reverberation-mapped AGN \citep{bentz13}.  Relatively
rapid variability of H$\beta$ would be predicted; that has never been
reported \citep[see also][]{veilleux16}.  Moreover,  the inner edge of
the circumbinary disk is estimated by \citet{yan15} to be 4.2 light
days from the center.  So, in the \citet{yan15} scenario, the BLR
would have to be located on top of the circumbinary disk, close to
its inner edge, and, as discussed above, would have to emit more than
100 times the normal flux of a typical broad-line region. 

\subsection{Comparison Sample}

In order to interpret the line emission from Mrk~231, we compiled the
characteristic properties of a small sample of Seyfert galaxies and
quasars for comparison. Infrared spectra of nearby Seyfert galaxies
and quasars were presented by \citet{landt08}, and we took
\ion{He}{1}*, P$\beta$, and P$\alpha$ (when available) line fluxes
from this reference.  We excluded Mrk~590 because it has turned into a
Seyfert 2 \citep{denney14}.  We excluded NGC~3227 as it is highly
reddened \citep{crenshaw01}.  Our sample included 15 objects with
measurements of these three lines and  \ion{C}{4} from the literature
(see below), and three more objects with measurements of
\ion{He}{1}*, P$\beta$, and \ion{C}{4} only. Among these 18 objects,
14 have black hole masses\footnote{http://www.astro.gsu.edu/AGNmass/}.
The range of log black hole masses represented, in units of solar
mass, is 6.88 to 8.84, with a mean of 7.66 and standard deviation of
0.54.  The range therefore is a bit higher than the log black hole
mass inferred for the small mass black hole (6.65 [solar
  masses]). However, we found that two objects with small log black
hole masses, NGC 4051 (6.130) and NGC 4748 (6.41), have infrared line
properties \citep{riffel06} consistent with the range of our
comparison sample.    

\citet{landt08} attempted to deconvolve the emission lines in terms of a
broad line and a narrow line.  This is an uncertain procedure unless
there is a break in slope between the broad and narrow line (i.e., as
in a Seyfert 1.5 or 1.8); in general, it is difficult to determine how
much of the line should be ascribed to the narrow line region,
especially when the line is cuspy.  To avoid this uncertainty, we used
the total line flux, i.e., the sum of 
the narrow and broad line fluxes in \citet{landt08} Table 5.  We note
that the narrow line flux is generally much smaller than the broad
line flux (the median ratio of narrow to broad line flux is about
9\%), so this approximation does not increase the uncertainty
significantly.  Moreover, we used the line ratios of the comparison
sample simply as an indication of the range of ratios observed in
nature, so high precision is not required.

Reddening can alter the line ratios, but to different degrees
depending on the ratio.  Reddening influences the
\ion{He}{1}*/P$\beta$ ratio minimally for typical objects; for
example, a modest $E(B-V)=0.1$ and an SMC reddening curve results in a
decrease in this ratio by about 2.3\%. The difference is larger for a
heavily-reddened object like Mrk~231, where the \citet{goobar08}
reddening curve with the parameters measured by \citet{leighly14}
yield a decrease in the ratio of 8.3\%.  Reddening is somewhat more
important for the P$\alpha$/P$\beta$, where $E(B-V)=0.1$ increases the
ratio by 3.4\% for SMC, and 19\% for the reddening measured in
\citet{leighly14}. Reddening is much more important for the
\ion{He}{1}*/\ion{C}{4} ratio, which increases by 290\% for
$E(B-V)=0.1$, i.e., a factor of 126 times larger effect than for the
\ion{He}{1}*/P$\beta$ ratio.  So, we think that much of the origin of
the large range of \ion{He}{1}*/P$\beta$ ratios is intrinsic, rather
than a consequence of reddening, because the dispersion, parameterized
by the standard deviation divided by the mean, is similar for the 
\ion{He}{1}*/P$\beta$ ratio (0.37) and the \ion{He}{1}/\ion{C}{4}
ratio (0.81). If reddening were dominating the range of observed
\ion{He}{1}*/P$\beta$ ratios, then a much larger dispersion would be
expected for the \ion{He}{1}/\ion{C}{4} ratio.  

We also included data from the intrinsically X-ray weak quasar PHL
1811 \citep{leighly01, leighly07, leighly07a}.  \citet{veilleux16}
draw comparisons with this object, positing that Mrk~231 has some
similarities with PHL~1811 analogs and weak-lined quasars.  We analyzed
a spectrum from our observations made using IRTF SpeX on 2008 August
22, 23, and 24.  We fit the continuum with a polynomial, and the
emission lines with Lorentzian profiles in the \ion{He}{1}* / P$\beta$
region, requiring that the Paschen lines
(i.e., the isolated P$\beta$ line and the blended P$\gamma$ line) have
the same width and a separation based on rest wavelengths.  Several
\ion{Fe}{2} lines were modeled with Gaussians. We model the
self-absorption on the \ion{He}{1}* emission line using a Gaussian
optical depth profile. P$\alpha$  is relatively isolated and was well
modeled with a Lorentzian profile.    

We obtained \ion{C}{4} measurements from the literature.  We dropped
IRAS~1750$+$508, H1934$-$063, and H~2106$-$099 from the sample because
they had had no UV spectroscopic observations. We extracted as many
measurements as possible from 
\citet{kura02}, \citet{kura04}, and/or \citet{tilton13} because they
presented a uniform analysis of large sets of {\it HST} spectra.
Other sources of {\it HST} measurements included \citet{obrien05} for
PDS~456,  
\citet{laor94} for 3C~273 and H1821$+$643,  \citet{kriss00} 
for NGC~7469, and \citet{leighly07a} for PHL~1811.  There were only
{\it   IUE} measurements available for Mrk~79, PG~0844$+$349, Mrk~110
and NGC~4593, compiled in \citet{wang98}.  These were used with
caution because comparison of values from the subset of objects that
also had observations using {\it HST} showed that the \citet{wang98}
measurements were consistently a factor of $\sim 10$ lower.  We
suspected that the flux-units footnote on \cite{wang98} Table 1 is too
low by a factor of 10, and we therefore multiplied their values by 10.  

\begin{figure*}[!t]
\begin{center}
\epsscale{1.0}
\includegraphics[width=4.5truein]{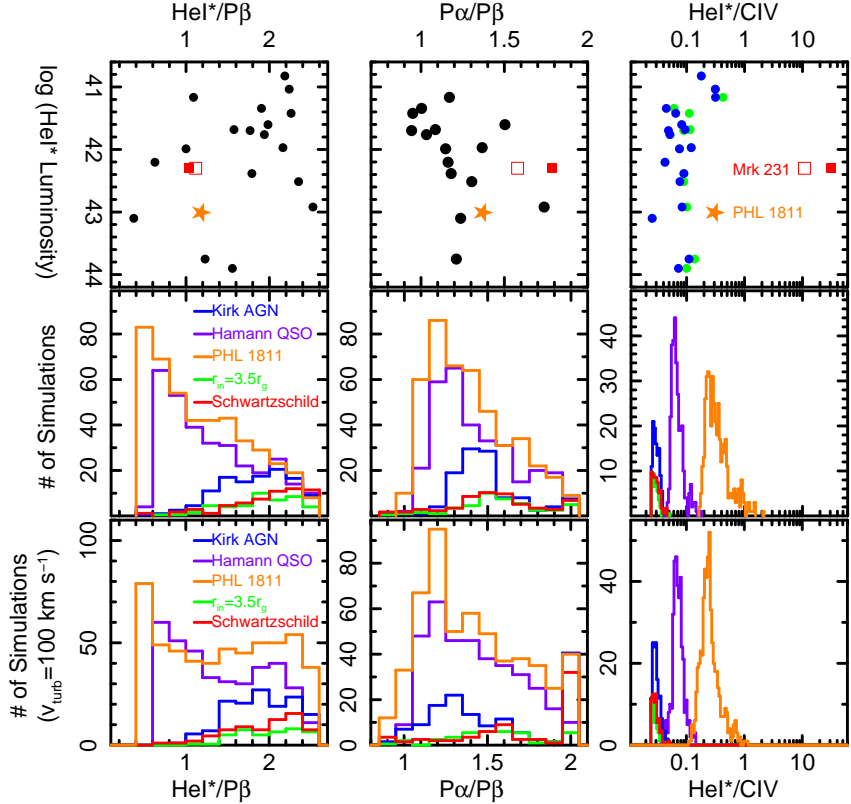}
\caption{{ Top Panels:}  Observed
  \ion{He}{1}*/P$\beta$, P$\alpha$/P$\beta$,  and
  \ion{He}{1}*/\ion{C}{4} ratios from Mrk~231 (filled red
  squares), the  intrinsically X-ray   weak quasar PHL 1811 (orange
  star), and a comparison sample   taken from \citet{landt08} (filled
  circles).  Also   shown are ratios for Mrk~231   corrected using the
  reddening curve   inferred by \citet{leighly14}   (for infrared ratios
  \ion{He}{1}*/P$\beta$ and P$\alpha$/P$\beta$)   and using an SMC
  reddening curve with $E(B-V)=0.1$ for the   \ion{He}{1}*/\ion{C}{4}
  ratio (open red squares).   Mrk~231 and PHL 
1811 have typical infrared line ratios, while  PHL~1811 has a somewhat 
high  \ion{He}{1}*/\ion{C}{4} ratio, due to its weak high-ionization
line emission \citep{leighly07a}.  The \ion{He}{1}*/\ion{C}{4} ratio
for Mrk~231 is much higher (note the logarithmic x-axis), due to the low 
\ion{C}{4} flux.  It is likely that we do not observed the intrinsic
\ion{C}{4} emission from Mrk~231 directly. { Middle Panels:} The 
distributions of results from  {\it Cloudy} simulations found to be
consistent with the adopted ranges of \ion{He}{1}*/P$\beta$ and 
P$\alpha$/P$\beta$ ratios, the lower limit of \ion{He}{1}*/\ion{C}{4}, 
and observed values of either \ion{He}{1}* flux or equivalent width.
These panels show that the {\it Cloudy} models are able to match the
observed ranges of the comparison sample ratios, and the
\ion{He}{1}*/P$\beta$ ratio suggests diagnostically useful SED
dependence, along with \ion{He}{1}*/\ion{C}{4}, assuming appropriate
reddening corrections can be made. { Bottom Panels:} The same as 
the middle panels, but including a turbulent velocity in the
simulations. The results are not substantially
different from the static case.  \label{fig3}}     
\end{center}
\end{figure*}

The UV spectroscopic observations were not made simultaneously with
the infrared spectroscopic observations, so relative variability is a
concern.   We estimated the importance of this effect by compiling
\ion{C}{4} measurements from multiple {\it HST} observations when
available: these are shown by blue and green points in
Fig.~\ref{fig3}. 
  Examination of this figure shows that the \ion{He}{1}*/\ion{C}{4}
ratio varies more across the sample than it does for any single
object (quantified below), and therefore relative variability is not
important.  Again, our intention was to estimate the range of the
\ion{He}{1}*/\ion{C}{4}  ratio observed in nature, and high precision
was not necessary. 

\subsection{Observed Ranges of Line Properties\label{ratios}}

We derived the range of line properties observed in the comparison
sample.  These ranges were used to compare with those from Mrk~231,
and to constrain the simulations discussed
in \S\ref{cloudy}.  Fig.~\ref{fig3} shows the ratios
\ion{He}{1}*/P$\beta$, P$\alpha$/P$\beta$, and \ion{He}{1}*/\ion{C}{4}
as a function of luminosity in the \ion{He}{1}* line.  

The \ion{He}{1}*/P$\beta$ ratio is shown in the top left
panel.  This ratio lies between 0.37 and 2.52 (a range of a factor of $\sim
7$), with mean and standard deviation of $1.67 \pm 0.61$.  For
Mrk~231, we show the observed ratio, and the ratio after correcting
for the reddening inferred in \citet{leighly14}.   Neither Mrk~231 nor
PHL~1811 have exceptional values of this ratio.  To constrain the
simulations, we use an upper limit of 2.5 (as observed) but extend the
lower limit to 0.1 in order to roughly compensate for possible 
contribution of \ion{He}{1}* emission from an outflowing component
that may not be present in P$\beta$.  

The P$\alpha$/P$\beta$ ratio is shown in the middle panel. This ratio
lies between 0.94 and 1.74 (a factor of 1.84), with mean and standard
deviation of $1.20 \pm 0.21$ for the comparison objects.  The observed
ratio for Mrk~231 is slightly high compared with this range (1.79);
however, when corrected for the reddening curve inferred by
\citet{leighly14}, the ratio drops to 1.58, a value roughly consistent
with that of the  comparison sample.  To constrain the simulations, we
consider values between 0.9 and 2.0.   

As discussed in \citet{of06}, in the low-density, low-optical-depth
limit (Case A), and the large-optical-depth limit whereby every
Lyman-line photon is scattered many times (Case B), 
P$\alpha$/P$\beta$ ratios of 2.33 and 2.28 are predicted, respectively.
Our ratios are significantly lower than that, suggesting high optical
depths in the broad line region.  

The range of observed \ion{He}{1}*/\ion{C}{4} ratios is large in the
comparison sample (0.025 to 0.37, a factor of $\sim 15$), with a mean
of 0.11  and a standard deviation of 0.089.  PHL~1811 has a rather
large ratio of  $\sim 0.3$.  The values for Mrk~231 are very large:
31.5 as observed, and 10.8 corrected for $E(B-V)=0.1$ using an SMC
reddening curve (see \S~\ref{sed}).  We note that it is not possible
to correct \ion{C}{4} from Mrk~231 using the \citet{goobar08}
reddening curve; as can be seen in Fig.~\ref{fig2},   the reddening is
completely optically thick at those wavelengths, and the observed
continuum and lines have a different origin, perhaps in scattered light
(\S \ref{spec}).   For simulations we set a lower limit of 0.025 on
the \ion{He}{1}*/\ion{C}{4} ratio, but do not set an upper limit in
order to see if we can attain the high values observed from Mrk~231.  

\section{Spectral Energy Distributions\label{sed}}

As discussed above, \citet{yan15} assumed a spectral energy
distribution that maximally favors their interpretation.    In this
section, we reconstruct their SED, correcting for the observed X-ray
emission, and compare it with more typical spectral energy
distributions that roughly span the range observed from AGN and
quasars.  We also consider  X-ray weak SED observed from PHL~1811.
The properties of the spectral energy  distributions are listed in
Table 1.  

\begin{figure*}[!t]
\begin{center}
\epsscale{1.0}
\includegraphics[width=4.0truein]{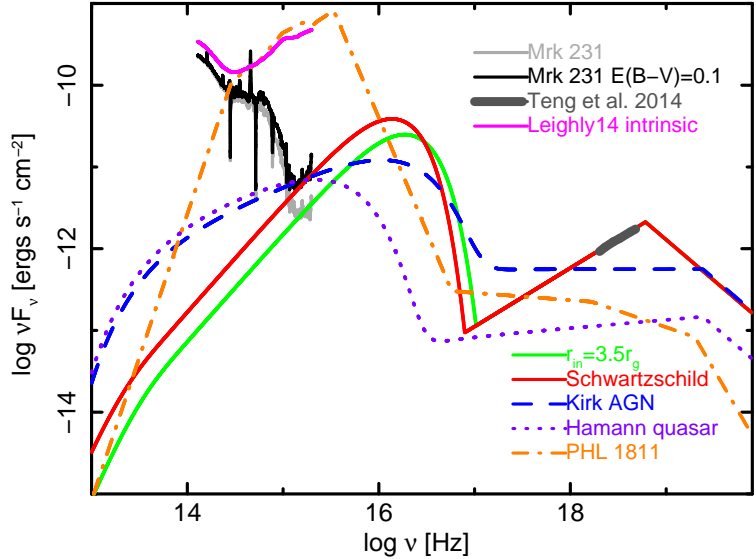}
\caption{Continua used for {\it Cloudy} simulations, overlaid on the
  merged   Mrk~231 near-UV-to-near-IR merged spectrum from
  \citet{leighly14}, and the X-ray results from
  \citet{teng14}. Representative   ``Kirk AGN'' and ``Hamann quasar''
  SEDs were used to demonstrate   that the {\it Cloudy} simulations
  can produce observed \ion{He}{1}*   equivalent widths, and
  \ion{He}{1}*/P$\beta$, P$\alpha$/P$\beta$, and
  \ion{He}{1}*/\ion{C}{4}   ratios in the comparison sample (\S
  \ref{comparison}).  The   ``$r_{in}=3.5 r_g$''   and
  ``Schwartzschild'' continua were used to 
  simulate the   \citet{yan15} model (\S \ref{yan_cloudy}). Finally,
  the ``PHL 1811''   continuum was   used to investigate the line
  fluxes and ratios for an X-ray weak   continuum (\S
  \ref{phl1811}).  \label{fig4}}   
\end{center}
\end{figure*}

We first considered two spectral energy distributions previously used to
model lines from AGN and quasars. A relatively hard one was adopted
from \citet{korista97}\footnote{This SED is taken as a typical AGN
  spectral energy distribution and is called by the {\it Cloudy}
  command {\tt AGN kirk}, or, equivalently {\tt AGN   6.00 -1.40 -0.50
    -1.0}.}, and a relatively soft one was taken from
\citet{hamann11}\footnote{This   SED is called by   the {\it Cloudy}
  command {\tt AGN T=200000K,     a(ox)=-1.7,     a(uv)=-0.5,
    a(x)=-0.9}.}.  These spectral energy   distributions 
are shown in Fig.~\ref{fig4}, arbitrarily normalized to   intersect the
near-UV portion of the SMC-$E(B-V)=0.1$-corrected Mrk~231 spectrum.    

\begin{deluxetable}{lccccc}
\tabletypesize{\scriptsize}
\tablewidth{0pt}
\tablecaption{Properties of Spectral Energy Distributions}
\tablehead{
\colhead{} & \multicolumn{2}{c}{Comparison Sample} &
  \multicolumn{2}{c}{Yan et al. 2015} & \colhead{Possible Intrinsic}  \\
\colhead{Property} & \colhead{Kirk AGN} & \colhead{Hamann QSO} &
\colhead{$r_{in} = 3.5 r_g$} & \colhead{Schwartzschild} & \colhead{PHL
  1811}}
\startdata
Reference & \citet{korista97} & \citet{hamann11} &
\multicolumn{2}{c}{\citet{yan15}$+$\citet{teng14}} & \citet{leighly07} \\
log Bolometric [$\rm erg\, s^{-1}$] & N/A & N/A &
44.34\tablenotemark{a} & 44.50\tablenotemark{b} &
45.70\tablenotemark{c} \\ 
$L/L_{edd}$ & N/A & N/A & 0.39\tablenotemark{d} & 
0.56\tablenotemark{d} & 0.17\tablenotemark{e} \\
$\alpha_{ox}$ & $-1.40$ & $-1.70$ & $-1.28$ & $-1.42$ & $-2.26$ \\
Log Q [$\rm photons\, s^{-1}$] & N/A & N/A & 54.36 &
54.61 & 55.61 \\
\enddata
\tablenotetext{a}{SED normalized to observed Mrk~231 continuum at
  2000\AA\/.  See Fig.~\ref{fig4} and text for details.}
\tablenotetext{b}{SED normalized to Mrk~231 continuum corrected for
  $E(B-V)$ at 2000\AA\/.  See Fig.~\ref{fig4} and text for details.}
\tablenotetext{c}{SED normalized to \citet{leighly14} inferred
  intrinsic Mrk~231 continuum at
  2000\AA\/.  See Fig.~\ref{fig4} and text for details.}
\tablenotetext{d}{Calculated using black hole mass $4.5 \times 10^6 \rm \, M_\odot$
  \citep{yan15}.}
\tablenotetext{e}{Calculated using black hole mass $2.3\times 10^8\rm \, M_\odot$ \citep{leighly14}.}
\label{table1}
\end{deluxetable}

Next, we reconstructed the \citet{yan15} SEDs.   \citet{yan15} 
infer a small amount of intrinsic reddening, between $E(B-V)=0.07$ and
$E(B-V)=0.14$, depending on the model, for an SMC reddening curve
\citep{pei92}.  Therefore, we required the spectral energy
distributions to intersect the near UV emission observed from Mrk~231
at 2000\AA\/, either as observed, or dereddened by an intermediate
value, $E(B-V)=0.1$.   

\citet{teng14} presented an analysis of {\it Chandra} and {\it NuSTAR}
observations of Mrk~231.    They found an absorbed, weak hard X-ray
continuum with a flat ($\Gamma\sim 1.4$) photon index that they take
to be the intrinsic X-ray continuum.   \citet{yan15} did
not consider the X-ray emission in their paper.   They felt that they
were justified in this assumption because \citet{teng14} noted that
Mrk~231 is X-ray weak ($\alpha_{ox}\sim -1.7).$\footnote{$\alpha_{ox}$
  is the point-to-point slope 
  between the ultraviolet continuum measured at 2500\AA\/ and the
  X-ray continuum measured at $2\rm \, keV$.}  However,
\citet{teng14} inferred that the observed UV is absorbed, and
estimated $\alpha_{ox}$ was based their estimate of the {\it intrinsic}
optical/infrared spectrum \citep[][Figure 3]{veilleux13}, not
the observed one.    With respect to the observed UV continuum, 
Mrk~231 is rather X-ray bright.   

The deconvolved X-ray spectrum is shown in Figure 5 of
\citet{teng14}.  We digitized the ``Direct PL'' component of that plot
between 8.4 and 19.9 keV.  We multiplied the result by a factor
of 1.29 in order to match the 0.5--30 keV luminosity in their Table 1
for the MyTorus model.  The \citet{teng14} power law is shown in dark
gray in Fig.~\ref{fig4}.   We required that the reconstructed \citet{yan15}
spectral energy distributions intersect this X-ray data, and have
$F_\nu \propto \nu^{-0.28}$ (the slope we measured from the digitized
spectrum), breaking to a slope of $-2$ for energies higher than 25
keV, the limit of the {\it NuSTAR} spectrum.     

\citet{yan15} assumed an inner radius $r_{in} = 3.5$ gravitational
radii ($r_g$) in order to
produce a radiative efficiency of $\eta = 0.1$.  Their
best-fitting model yielded  $M_{\odot} \sim 4.5\times 10^6 \rm \,
M_\odot$ radiating at $0.6 L_{Edd}$ for the smaller-mass black hole.
They  assumed that the optical-UV spectrum of the smaller black hole is
characterized by a sum-of-blackbodies accretion disk model
\citep[e.g.,][]{fkr02}.  The outer edge of the disk was taken to be
100 times the inner edge.  This information was sufficient for us to
follow \citet{yan15} and compute a disk spectrum from the infrared
through the far UV  (i.e., through the high-temperature rolloff).  The
disk spectrum was normalized to the observed 2000\AA\/ flux, and the
X-ray spectrum as described above joined to the infrared-through-UV
spectrum.  This SED is shown in Fig.~\ref{fig4}, and information about
this continuum is given in Table 1.  

When we normalized the $r_{in} = 3.5 r_g$ spectrum to the Mrk~231
continuum corrected for $E(B-V)=0.1$ at 2000\AA\/, we found that the 
emission is super-Eddington ($L/L_{Edd}=1.14$).  \citet{yan15} found
it to be sub-Eddington, possibly  because they did not include the
X-ray emission.  So, we normalized the $r_{in}=3.5 r_g$ SED to the 
observed Mrk~231 continuum at 2000\AA\/, rather than the $E(B-V)$
dereddened one.  But in order to give the \citet{yan15} model the best
chance for success (i.e., the largest possible photoionizing flux), we
also considered a Schwartzschild disk, i.e., $r_{in} = 6 r_s$,
normalized to the Mrk~231 spectrum corrected for intrinsic absorption
of $E(B-V)=0.1$; it is also shown in Fig.~\ref{fig4}. This SED is not 
super-Eddington (Table 1).     

These two spectral energy distributions have relatively flat values
of $\alpha_{ox}$ of  $-1.28$, when normalized to the observed
continuum, and $ -1.45$ when normalized to the $E(B-V)=0.1$ dereddened
one. $\alpha_{ox}$ is related to the UV monochromatic 
luminosity at 2500\AA\/ \citep[e.g.,][]{steffen06}; those regression
relationships predict $\alpha_{ox}$ to be $\sim -1.6$.  This is
steeper than inferred, but roughly within the regression
uncertainty of $0.24$ \citep[][Equation 2]{steffen06}.  Since,
according to the \citet{yan15} model, the bulk of the X-ray emission
emerges from the central engine of the small-mass black hole, that
system should be relatively X-ray bright, like a Seyfert nucleus.   

Fig.~\ref{fig2} and Fig.~\ref{fig4} display the intrinsic continuum
inferred using the circumstellar reddening model \citep{leighly14}.
When we used our inferred intrinsic continuum and the extrapolation of
the \citet{teng14} power law, we obtained $\alpha_{ox}$ of $-2.2$,
essentially the same value observed from the intrinsically X-ray weak
quasar PHL~1811 \citep[between   $-2.2$ and   $-2.4$, accounting for
  X-ray   variability,][]{leighly07}.  {   Moreover,
  \citet{veilleux16} also   note the similarity between   properties
  of Mrk~231 and PHL~1811 and    its analogs.}  Intrigued by this
result, we investigated whether the extreme X-ray weak PHL~1811
continuum could produce the emission lines observed in Mrk~231 in
\S\ref{phl1811}.

\section{Cloudy Models\label{cloudy}}

We  used the photoionization code {\it Cloudy} \citep{ferland13} to
see if we could produce the \ion{He}{1}*, P$\beta$, P$\alpha$, and
\ion{C}{4} lines in the range of strengths and ratios observed.  We
performed a set of simulations for each of the five continua described 
in \S\ref{sed}.  In each case, we perform 5000 or 10,000 simulations
with parameters randomly drawn from uniform distributions of
ionization parameter ($ -3 \leq \log(U) \leq 1.0$), density ($6 \leq
\log(n) \leq 11.5$), and a combination parameter defined as the
difference between the log of the column density $N_H$ and the log of
the ionization parameter, $\log(N_H)-\log(U)$, that measures the
thickness of the gas slab relative to the hydrogen ionization front
($22.5 \leq \log(N_H)-\log(U) \leq 24.0$), and has been shown to be
useful in analysis of both emission lines and absorption lines
\citep{leighly04,   casebeer06, leighly07a,   leighly09, leighly11,
  leighly14, lucy14}.  We performed the entire set of simulations for
a stationary gas, and for a gas with a turbulent velocity
$v_{turb}=100\rm \, km\, s^{-1}$.  A total of 80,000 simulations were
performed.     

To characterize the strength of the \ion{He}{1}* emission, we used
either the line flux or the line equivalent width, depending on the
circumstance.   For the \citet{yan15} models, we use the flux of the
line, since, as discussed in \S\ref{mrk231}, huge equivalent widths
with respect to the photoionizing continuum are required. For the 
comparison objects, we used the line equivalent width, although
comparison of theoretical and observed equivalent widths can be
difficult for infrared lines. For example, 
the {\it   Cloudy} line fluxes assume full coverage, yet it is known
that the broad line region does not fully cover the continuum (or we
would always see absorbed continuum spectra in the X-ray 
band). \citet{leighly04} found a covering fraction of 0.05 for
intermediate- and low-ionization lines for two rather weak-lined
quasars, and we used that value for the lower limit.  For the upper
limit, we chose a value of 0.5.    

Also, the equivalent width from simulations will be with respect to
the AGN continuum, while the observations may include a torus
component in the near infrared, and/or host galaxy.  The comparison
sample is dominated by nearby AGN and quasars, and it is likely that
the spectroscopic slit excluded much of the galaxy contribution.  The
\ion{He}{1}* line, at 10830\AA\/, occurs just at the 1-micron break,
where the accretion disk continuum and the torus contribution are
approximately equal, and so it is expected that the torus contribution
will not dominate. 

In addition, the near-infrared portion of the continuum, near 1
micron, lies far from the hydrogen continuum shortward of 911\AA\/,
which sets the level of the photoionizing flux. For example, the two
SEDs used in \S\ref{comparison} have optical-UV spectra with $F_\nu
\propto \nu^{-0.5}$, typical of quasars and AGN
\citep[e.g.,][]{natali98}.  But the intrinsic slope is not uniform
among AGN, and the larger the difference from $\nu^{-0.5}$, the larger 
the incurred uncertainty in the equivalent width in the infrared band.  

We addressed these concerns by considering a large range of
equivalent width for \ion{He}{1}*.  The observed equivalent width
range in the comparison sample is 30--300\AA\/, and therefore we
accept simulations producing equivalent widths for full covering
between 60\AA\/ (i.e., lower limit on equivalent width divided by
upper limit on covering fraction) and 6000\AA\/ (i.e., upper limit on
equivalent width divided by lower limit on covering fraction). 

Generally speaking, we chose very generous bounds on every parameter
in order to give the \citet{yan15} model the best chance.  Therefore,
when we show it is not feasible, our result 
cannot be attributed to an artificial or arbitrary limitation to the
considered range of parameter space.   

\subsection{Cloudy Models of the Comparison Sample\label{comparison}}

We first needed to establish that {\it Cloudy} can explain the observed
\ion{He}{1}* intensity and the observed line ratios from typical
objects  to be confident that we could use the {\it 
  Cloudy} results to analyze special cases.  We bracket the plausible
range of SED shapes by considering a hard, X-ray bright one from
\citep{korista97} that may be appropriate for Seyfert galaxies, and a
soft one that may be appropriate for QSOs \citep{hamann11}.       We
extracted the predicted \ion{He}{1}*, P$\beta$, 
P$\alpha$, and \ion{C}{4} fluxes from the simulation results using
these two SEDs, and computed the \ion{He}{1}*/P$\beta$,
P$\alpha$/P$\beta$, and \ion{He}{1}*/\ion{C}{4} ratios.    

We accepted solutions that were consistent with the line ratios in the
ranges discussed in \S\ref{ratios}:  \ion{He}{1}*/P$\beta$ between 0.1
and 2.5, P$\alpha$/P$\beta$ between 0.9 and 2.0, and
\ion{He}{1}*/\ion{C}{4} larger than 0.025.  For the \ion{He}{1}* flux
constraint, we used the equivalent width range (for full covering)
between 60 and 6000\AA\/, as discussed in \S\ref{cloudy}.  

We show the histograms of results from accepted simulations in the
lower panels  of Fig.~\ref{fig3}.  Interestingly, SED dependence is present
in all the ratios, to a greater or lesser degree.  The soft continuum
leans towards smaller values of \ion{He}{1}*/P$\beta$, while the hard
continuum yields a larger value of this ratio. This is plausibly a
consequence of the the stronger helium continuum in the hard SED.
Both spectral energy distributions favor an intermediate value of
P$\alpha$/P$\beta$, interestingly close to the mean value observed.
This ratio depends principally on optical depth, so a great deal of
SED dependence is not expected.   

The two SEDs produced the largest differences in the
\ion{He}{1}*/\ion{C}{4} ratio.  The hard SED produces low values 
of this ratio; that is, it yields relatively large \ion{C}{4} fluxes.
This is expected; a harder SED will produce a hotter photoionized gas
\citep[e.g.,][Fig.\ 14]{leighly07a}, and \ion{C}{4} is an important
coolant, so the proportion of \ion{C}{4} compared with a recombination
line like \ion{He}{1}* can be expected to be large when the SED is
hard.   The softer SED, yields a lower value and  matches the observed  
distribution of \ion{He}{1}*/\ion{C}{4} ratios  nicely.  

\begin{figure*}[!t]
\begin{center}
\epsscale{1.0}
\includegraphics[width=4.5truein]{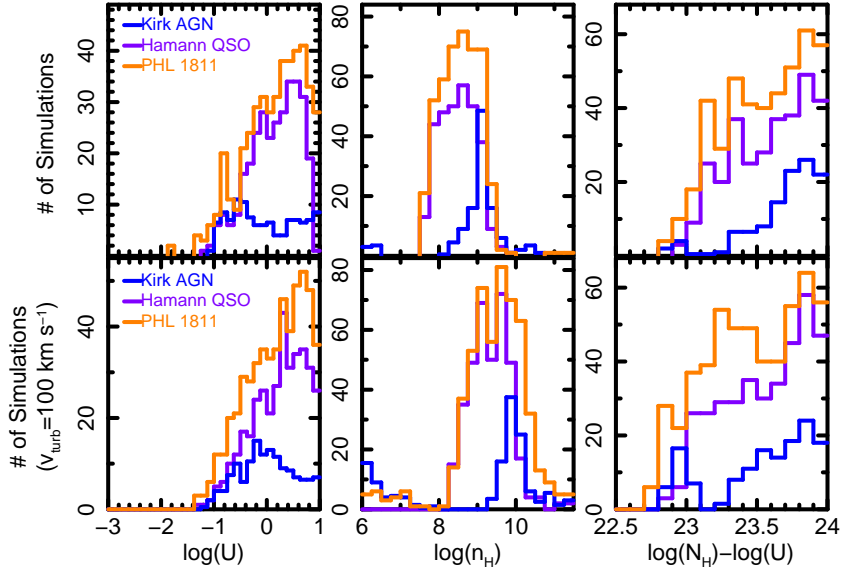}
\caption{The  distributions of input parameters from
  {\it     Cloudy} simulations in which the results were selected to
  be consistent with the   observed   ranges of
  \ion{He}{1}*/P$\beta$  and P$\alpha$/P$\beta$   ratios,   and the
  lower limit of    \ion{He}{1}*/\ion{C}{4} ratios in the comparison
  sample, and  either   \ion{He}{1}* flux or equivalent width; 
see text for details.    We found that the
simulations are consistent   with a rather high ionization parameter
(higher for softer SEDs, as   expected, to meet the requirement of a
sufficiently strong helium   continuum to yield the observed
\ion{He}{1}* emission), an   intermediate range of density, and a
relatively high column   density, i.e., $\log(N_H) -\log(U) \ge \sim
23$.  The top panels   show the results for the stationary case, and
the bottom panels   show the results for $v_{turb}=100 \rm \, km\,
s^{-1}$.  The   principal difference is the favored density range,
which   is shifted to higher densities for the turbulent case because
of reduced thermalization of the Paschen lines as a consequence of
lower opacity.   \label{fig5}}  
\end{center}
\end{figure*}

Fig.~\ref{fig5} shows the {\it Cloudy} input parameters for the
accepted simulations.  The softer SED tends to favor a higher
ionization parameter than the harder one.  This is expected; a higher
photon flux is necessary for a soft SED to produce the required
\ion{He}{1}* flux or equivalent width.

There is a strong localization in densities favored due to opacity of
the Paschen lines.  The simulation results show that for fixed $\log
U$ and $\log N_H - \log U$,  both P$\alpha$ and P$\beta$ became
thermalized at larger densities (i.e., the increase of line flux with 
density broke to a flatter slope), but with P$\alpha$ becoming
thermalized at slightly lower densities than P$\beta$, resulting in a
decrease in  the P$\alpha$/P$\beta$ ratio to less than the observed upper
limit of $\sim 2$ near $\log n_e = 7.5$, and thus providing a
constraint on the density on the low end.    At the same time,
P$\beta$ became thermalized much faster than \ion{He}{1}*, resulting
in a ratio higher than the observed upper limit for the
\ion{He}{1}*/P$\beta$ ratio of $\sim 2.5$ for values greater than
$\log n_e = 9.5$ and thus constraining the density on the high end. 

To investigate the influence of optical depth, we ran the
simulations including a turbulent velocity $v_{turb}=100\rm \, km\,
s^{-1}$.  The effect of turbulence is to decrease optical depths
\citep[e.g.,][]{bottorff00}.  The chosen value of $v_{turb}$ 
is much larger  than the  thermal line width in a photoionized gas
(about $15\rm \, km\, s^{-1}$).  Physically, it may represent actual
macro-turbulence, or differential velocity.  This value was chosen
arbitrarily because it was large enough to show an effect (no
significant effect was observed for $v_{turb}=15\rm \, km\, s^{-1}$),
and a single value allowed us to understand the effect of
turbulence qualitatively.   The principal effect of turbulence was a
shift of the favored range of density.  As expected, the opacity is
lower in the turbulent case, and thermalization becomes important at
higher densities than in the stationary case. 

There were a greater number of simulations accepted for larger values
of $\log(N_H)-\log(U)$, i.e., column density, as expected, since the
Paschen lines are produced predominantly in the partially ionized
zone, located beyond the hydrogen ionization front, i.e.,
$\log(N_H)-\log(U) \gtrsim 23.2$.     

These simulations show that these typical AGN and quasar SEDs are able
to produce the \ion{He}{1}* equivalent widths, and
\ion{He}{1}*/P$\beta$, P$\alpha$/P$\beta$,  and
\ion{He}{1}*/\ion{C}{4} ratios observed from the typical objects in
the comparison sample.  We now turn to the more specialized cases. 

\subsection{Cloudy Models using the \citet{yan15} Spectral Energy
  Distributions\label{yan_cloudy}} 

{\it Cloudy} models employing the $r_{in}=3.5r_g$ and the
Schwartzschild spectral energy distributions were used to evaluate
the viability of the \citet{yan15} model.   We first asked a basic 
question: can these SEDs, normalized as described to the observed UV
and X-ray emission, produce the \ion{He}{1}* flux observed, given the
range of observed \ion{He}{1}*/P$\beta$ ratios? 
The advantage of this minimal set of constraints is that it uses only
infrared spectral data, and lines that are close to one another, and so will be
minimally impacted by reddening regardless of model.  {\it We found 
  that neither of these two SEDs could produce the observed
  \ion{He}{1}* flux when   constrained to produce the observed range
  \ion{He}{1}*/P$\beta$ ratios,  even for a covering fraction of the
  broad line region equal to 1.} This result is not unexpected given
the discussion in \S~\ref{mrk231}, i.e., that the near-IR  equivalent
widths would have to be $\sim 100$ larger than normal with respect to
the photoionizing continuum in order to explain the near-infrared line
emission observed.  It is simply not reasonable to expect the
continuum from a $4.5 \times 10^6 \rm \, M_\odot$ black hole to be
able to power the broad line region emission of a $> 10^8 
\rm \,  M_\odot$ quasar.     

{  As noted in \S 2, \citet{yan15} state, in their \S 6, that there 
are sufficient photons in the small-black-hole-mass continua to
explain the H$\beta$ emission.  We believe that they underestimated
the required photon flux by using H$\beta$, since the spectrum is
significantly redddened in that region.  H$\beta$ is strongly blended
with the strong \ion{Fe}{2}, making it difficult to measure, but the
spectral fit shown in \citet{leighly14} Fig.\ 6 yielded an estimate
of the observed flux in H$\beta$ of $2.8 \times 10^{-13}\rm \, erg\,
s^{-1}$\AA\/$^{-1}$, with the dereddened value being 5.4 times larger.
The observed photon flux in H$\beta$ is then $2.6 \times 10^{53}\rm \,
photons\, s^{-1}$.  For  a temperature of $10^4\rm \, K$ and a density
of $10^6 \rm cm^{-3}$, and assuming Case B \citep{of06}, we find that
the photonionizing photon flux must be a factor of $\alpha_B /
\alpha^{eff}_{P\alpha} = 8.44$ times larger than the photon flux in
the line, i.e., $Q=2.2\times 10^{54}\rm \, photons \, s^{-1}$.  This
calculation assumes that the covering fraction is 100\%; for a more
realistic covering fraction, the photon flux would have to be even
larger.  The photoionizing fluxes for the \citet{yan15} continua are
given in Table\ref{table1}.  The photoionizing flux from the $3.5 r_g$ 
continuum just matches the required value (so full coverage is
necessary), and the value from the Schwartzschild continuum exceeds
the required value by a factor of 1.8, requiring a covering fraction
of $\sim 50$\%.  However, the dereddened H$\beta$ requires $Q=1.2
\times 10^{55}\rm \, photons \, s^{-1}$, exceeding the \citet{yan15}
ionizing photon fluxes by factors of 3--5.

P$\alpha$ is subject to less reddening, and may provide a more
accurate estimate of the required photoionizing flux.  The flux of
P$\alpha$ was measured to be $8.1\times 10^{-13}\rm \, erg\, s^{-1}\,
cm^{-2}$, for a photon flux of $2.9 \times 10^{54} \rm\, photons\,
s^{-1}$ in P$\alpha$.  The $\alpha_B / \alpha^{eff}_{P\alpha} = 6.99$
for this line implies that $Q = 2.0 \times 10^{55}\rm \, photons\,
s^{-1}$ for full coverage.  This value exceeds the photoionizing flux
from the Yan et al.\ continua by a factor of 5--9.  These values are
roughly consistent with the above estimate from the unreddened
H$\beta$ above.  Thus, we conclude that \citet{yan15} underestimated
the required photon flux because they used the significantly reddened
H$\beta$ line. 

Moreover, while adequate for planetary nebula and \ion{H}{2}
regions, the nebular approximation is well known not to be appropriate
for AGN \citep[e.g.,][]{dn79}, where the bulk of the hydrogen line
emission  arises from the partially ionized zone.  Our more complete
treatment uses infrared lines that are less likely to be affected by 
reddening, and requires that we only consider the simulations that yield
\ion{He}{1}*/P$\beta$ ratios within the range observed from AGN.
Fig.~\ref{fig5} shows that most of the   one-zone models that meet
this requirement lie in the partially ionized zone (with values of
$\log N_h - \log U > \sim 23.2$), as expected. }

We were also interested in whether these SEDs could explain
the unusually large \ion{He}{1}*/\ion{C}{4} ratio observed. The
equivalent width uncertainty outlined in \S\ref{cloudy} is exacerbated 
by the fact that the sum-of-blackbodies accretion disk spectrum has a
long wavelength spectrum described by $F_{\nu} \propto \nu^{1/3}$.
Such a steep   spectrum is not generally  seen in AGN; this is one of
the   problems hampering our    understanding of accretion disks
\citep[e.g.,][]{kb99}. Typical values of the optical to UV slope are
observed to be around $-0.5$ \citep[e.g.,][]{natali98,  vandenberk01}
to $-0.3$ \citep[e.g.,][]{francis91, selsing15}.  For SEDs with the
same value of $F_\lambda$ at 911\AA\/, the continuum will be 7.87
times weaker at 10830\AA\/ for a $F_\nu \propto \nu^{1/3}$ continuum
than for a $F_\nu \propto \nu^{-0.5}$ continuum.  To compensate for
this additional uncertainty, we increase the upper limit on the
allowed equivalent widths by a factor of 7.87.   

The results are shown in Fig.~\ref{fig3}.  The $r_{in}=3.5r_g$ and
Schwartzschild SEDs produce results that are similar to those
obtained with the harder SED explored in \S\ref{comparison}.  This is not
surprising given the large fraction of photoionizing flux in the
extreme UV and the flat values of $\alpha_{ox}$.  Specifically, they
predicted only very low values of \ion{He}{1}*/\ion{C}{4}, and cannot
explain the high value observed in Mrk~231.  

\subsection{Cloudy Models using the PHL 1811 Continuum\label{phl1811}}

As discussed in \S~\ref{sed}, the PHL~1811 continuum may be similar to
the intrinsic continuum in Mrk~231.  In this section, we explore
whether it can produce the observed emission lines.  We normalized the
SED to the \citet{leighly14} intrinsic continuum, and required the
simulations to produce the observed \ion{He}{1}* flux.  We first
deredden the \ion{He}{1}* flux using the inferred circumstellar
reddening curve from \citet{leighly14}.   A large fraction of the
simulations are able to meet these selection criteria, and their
properties are shown in Fig.~\ref{fig3} and Fig.~\ref{fig5}.     

Fig.~\ref{fig3} shows that the \ion{He}{1}*/P$\beta$, and P$\alpha$/P$\beta$
ratios strongly resemble those produced by the soft SED considered in
\S~\ref{comparison}, but the \ion{He}{1}*/\ion{C}{4} ratio is higher
than for the other SEDs, and is consistent with the observed value
from PHL~1811.   This is not surprising as the very X-ray weak
PHL~1811 SED has been shown to produce weak high-ionization lines
\citep{leighly07a}.  Yet the \ion{He}{1}*/\ion{C}{4} ratio does not
approach the very high value exhibited by Mrk~231.  As PHL~1811 is
intrinsically exceptionally X-ray weak \citep{leighly07}, with
exceptionally small \ion{C}{4} equivalent width \citep{leighly07a}, we
suspect that it would be difficult to produce a much higher ratio
intrinsically. This is evidence that  the \ion{He}{1}*/\ion{C}{4}
ratio in Mrk~231 is high because of reddening and is not intrinsic.
The UV continuum and broad-line region in Mrk~231 are likely not seen
directly at all.     

\section{Evidence for Resolved Far UV Emission\label{fuv}}
If the weak far-UV emission is not the continuum from the
smaller-black hole in a milli-parsec black hole binary system, then
what is it? In this section, we report the discovery of evidence for
resolved FUV emission in the archival {\it HST} STIS data of Mrk~231
that is supported by analysis of an {\it HST} FOC image.  This result
suggests that the far-UV continuum does not originate in the central
engine.  

\subsection{Spatial Analysis of the {\it HST} STIS Observation\label{stis}}

\citet{leighly14} reported analysis of the APO TripleSpec observation
of Mrk~231.  We found that the spectral trace was broader in the 
\ion{He}{1}*10830 trough than it was at unabsorbed wavelengths.  
This result indicated that the trough is partially filled in by
extended emission.  Given the relatively poor spatial resolution
available in the ground-based observation, and the infrared
bandpass, the extended emission is probably the host galaxy.

\begin{figure*}[!t]
\epsscale{1.0}
\begin{center}
\includegraphics[width=6.5truein]{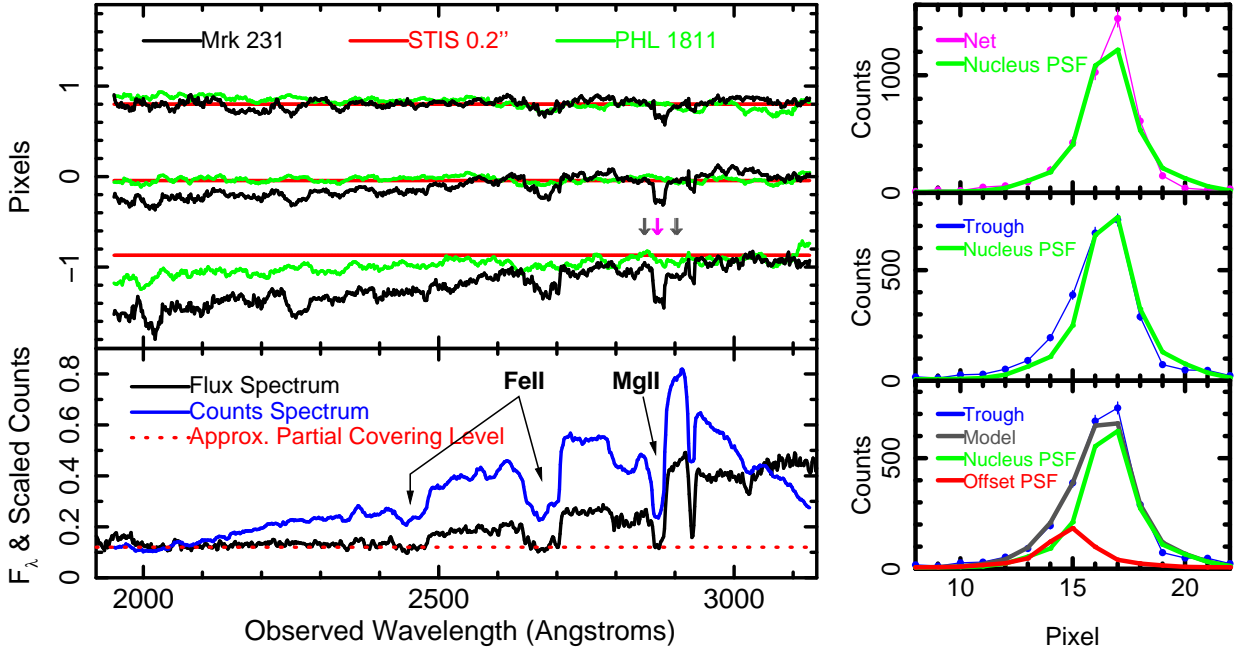}
\caption{Analysis of the spatial profiles of the {\it HST} STIS G230L
  observation of Mrk~231 indicating evidence for resolved emission in
  the far UV.  { Left:}  The lower panel shows the
  flux and counts spectrum.  An approximate constant extrapolation of
  the far UV continuum to longer wavelengths (red dashed line)
  suggests that the 
  \ion{Fe}{2} and \ion{Mg}{2} troughs are saturated and filled in by
  the same continuum.  The upper panel shows the analysis of the
  spatial profiles.  The 0.25, 0.5, and 0.75 levels (bottom, middle,
  and top traces) of the normalized  cumulative histogram are shown
  for the  STIS G230L point spread function for a 0.2 arcsecond slit at
  2400\AA\/ (red line), the comparison point source PHL~1811 (green
  line), and   Mrk~231 (black line).  An 
  offset of the 0.25 level is seen for Mrk~231, indicating extended
  emission, especially at short wavelengths, and in the \ion{Mg}{2}
  trough near 2873\AA\/ and   \ion{Fe}{2}   around 2680\AA\/.  {
    Right:}  Analysis of profiles 
  accumulated   near the magenta arrow in the left panel (the trough)
  and the   difference between the mean of the gray arrows and the
  trough (the   net).  The net profile is fit reasonably well by the
  STIS PSF   (top), but the trough exhibits significant left-side
  residuals   when fit    with the same PSF (middle).  A good fit is
  obtained with an   additional component offset by $1.72\pm 0.12$
  pixels, corresponding   to $36.5\pm 2.5\rm \, pc$.   \label{fig6}}   
\end{center}
\end{figure*}

The observed-frame {\it HST} STIS spectrum is shown in the lower panel
of Fig.\ref{fig6} \citep[see   also][Fig.\ 3]{veilleux16}. The
\ion{Mg}{2} trough observed near 2870\AA\/, and the
low-excitation \ion{Fe}{2} troughs near 2450\AA\/ and 2674\AA\/ 
(observed wavelengths) are approximately flat and have approximately
the same flux level as the far UV continuum.  This suggests that those
troughs are saturated and suffer partial covering, and the
far UV continuum partially fills in the troughs.  It is therefore
conceivable that interesting constraints on the origin of the far-UV
emission might be obtained from performing the same type of analysis
on the STIS data as was performed on the APO TripleSpec data.

Analysis of the spatial profile of the STIS is not trivial.  For
example, the detector under samples the point spread function;  the
STIS line spread function for the G230L detector at  
2400\AA\/\footnote{http://www.stsci.edu/hst/stis/performance/spectral\_resolution/LSF\_G230L\_2400\_dat}
has a FWHM of 1.67 pixels.  As noted in the STIS Data
Handbook\footnote{http://www.stsci.edu/hst/stis/documents/handbooks/currentDHB/ch5\_stis\_analysis5.html\#418394},
this property affects the rectified 2D images.  Specifically, the
spectra extracted from a single row of the rectified 2D images
produced by the standard pipeline processing include artifacts
(scalloping) due to interpolation from one row to the next.  Improved
rectification can be achieved by using {\tt wx2d}, a program available
in the IRAF package STSDAS.  This program employs a wavelet interpolation
for improved rectification \citep{bd06}.  We used this program on the
four STIS G230L unrectified images, and then added them together.  For
comparison, one observation of the bright $z=0.192$ quasar PHL~1811
was analyzed in the same way.

Spatial profiles were constructed for each wavelength bin between 1950
and 3127 \AA\/ (observed frame, to facilitate comparison with PHL
1811).  The signal-to-noise ratio for a single wavelength bin was low,
 so the median among the central wavelength bin and
a set number of wavelength bins to each side was used.  The
signal-to-noise ratio was poorest at shorter wavelengths, so the
number of bins on each side was chosen to be 10 and 5 for wavelengths
shorter and longer than 2600\AA\/ respectively.   The
nearly-negligible background was estimated to be the median of the 10 
pixels on each side of the central 13 pixels.  The  central 13 pixels
of the background-subtracted profile were oversampled by a factor of
10, and the cumulative histogram was compiled and normalized.
Finally, the pixel locations of the 0.25, 0.5, and 0.75 levels of the
cumulative histogram were identified by interpolation. The robustness
of this procedure was confirmed by varying various
parameters in the analysis, including the size of the central region,
the size of the background region, and the number of pixels used to
construct a profile.

The results are shown in Fig.~\ref{fig6}, left side.  The top panel
shows the distance between the pixel location of the 0.25, 0.5, and
0.75 levels of the cumulative histogram and the fitted center of the
mean profile of the whole image.  Superimposed are results obtained
by performing the same analysis on the spatial profiles of the bright
quasar PHL~1811, and on the STIS line spread function profile for
the G230L grating at 2400\AA\/ using a 0.2 arcsecond slit.  PHL
1811 is a bright, unresolved point source, so ideally, the pixel
locations of the 0.25, 0.5, and 0.75 levels should coincide with
those from the STIS  line spread function.  The correspondence is good
at long wavelengths, but less so at shorter wavelengths, where the
0.25 level location for PHL~1811 sags below the STIS line spread
function result by $\sim 0.2$ pixel.  This indicates that the spatial
profile for PHL~1811 is slightly broader than the STIS line spread
function. Part  of the difference could be due to the fact that at
shorter wavelengths the STIS line spread function is broader with more
prominent wings. The remaining difference is taken to represent
the residual systematic uncertainty in the rectification of the 2D
images.  

The Mrk 231 profile differs from both the STIS line spread function
and the PHL 1811 profile.  While the pixel location of
the 0.75 cumulative histogram level coincides with that of the STIS
line spread function and the PHL~1811, the pixel locations of the 0.5
and especially the 0.25 levels are offset.  This indicates that the
spatial profile of Mrk~231 is quite a bit broader than that expected 
from a point source, especially at short wavelengths, as well as being
asymmetric.  The count rate is low at short wavelengths, so one might
be tempted attribute this offset as systematic uncertainty in the in
the rectification of the 2D image.  However, {\it the asymmetry is also
  observed in the absorption line troughs}.   This is most  
clearly seen in the \ion{Mg}{2} trough observed around 2873 \AA\/, and
the low-excitation \ion{Fe}{2} trough observed around 2680 \AA\/, but
is also detectable to a lesser extent in the \ion{Mg}{1} trough
observed around 2930\AA\/.  This is precisely the behavior expected if 
the far-UV continuum is extended and fills in the bottoms of
saturated troughs.  

What is the physical extent of the asymmetry?  We compiled profiles  
characterizing the \ion{Mg}{2} trough, centered around 2870\AA\/, and
marked on Fig.~\ref{fig6} by a magenta arrow, and total NUV continuum
emission, taken to be the mean of the profiles centered around 2848 and
2902\AA\/, marked on Fig.~\ref{fig6} by gray arrows.  (We used the mean
of two points bracketing the \ion{Mg}{2} trough to approximately
compensate for the monotonic decrease of the grating effective area at
these wavelengths).  At each point, we constructed the profile using
the sum of the central pixel and 5 pixels on each side to increase the
signal-to-noise ratio.  As the STIS MAMA is a photon counting
detector, we assume that the uncertainty on these counts profiles
is Poisson.     

If the \ion{Mg}{2} trough is filled in by the far UV continuum, the
difference between the total continuum and the trough continuum
profiles, referred to as the net continuum profile, can be assumed to
characterize the profile of the NUV point source, in particular, its
spatial location.  We fit this profile using CIAO
Sherpa\footnote{http://cxc.harvard.edu/sherpa4.8/} with a template
model created from the STIS G230L line spread function a range of pixel
offsets.   to determine the pixel location of the nucleus.  The best
fit to the net profile locates the pixel location in the spatial
direction of the nucleus on the 2-D spectral image (top right panel of
Fig.~\ref{fig6}). 

If Mrk 231 were consistent with a point source at every wavelength,
then spatial profiles compiled at any wavelength should be well fit by
the STIS G230L spatial profile with the same offset as the net
profile.  The second panel in Fig.~\ref{fig6} shows the trough profile
subject to such a fit.  A significant wing is observed on the left
side of the profile, invalidating the assumption that Mrk~231 FUV
emission is consistent with the nuclear point source, and indicating
the presence of extended emission. 

In order to quantify the extent of the extended emission, we make the 
simple assumption that the Mrk~231 FUV profile consists of the
nuclear emission plus another, offset component.  We fit the trough 
profile with the net profile model component, and a second offset
profile.  The result is shown in the lowest right panel of
Fig.~\ref{fig6}.  This model provides a good fit to the wing on the
left side of the profile.  The offset component accounts for 20\% of
the total flux, and the  separation is  $1.73\pm 0.12$ pixels,
corresponding to $0.0424 \pm 0.0029$ arcseconds.  At the distance of
Mrk~231, 1 arcsecond corresponds to 863 parsecs \citep{veilleux13}, so
the offset is $36.5 \pm 2.5\rm \, pc$.  We note that this is a lower
limit on the offset, since the slit width was 0.2 arcseconds (i.e.,
8.16 pixels) and we don't know the relative orientation of the slit
and the offset emission. 

\subsection{Spatial Analysis of the {\it HST} FOC F210M Image}

The analysis presented in \S\ref{stis} suggests the presence of far-UV
resolved emission on the scale of 0.042 arcseconds.  The
diffraction-limited resolution for 
{\it HST} at 2000\AA\/ is 0.017 arcseconds, so in principle the 
resolved emission could be detected in an image.  The archive contains
only one imaging observation in the far UV, a 596.5-second observation
taken 1998 Nov 28 using the Faint Object Camera (FOC) with the F210M
filter (pivot wavelength = 2180\AA\/, 162\AA\/ RMS bandwidth) as
part of the imaging polarimetry campaign on Mrk 231. The plate
scale for the f/96 relay is $0.01435 \pm 0.00007$'' per FOC pixel.
The polarimetry  observations were made using F346M, with central
wavelength 3400\AA\/, and are therefore dominated by the unresolved 
near-UV/optical component, and so can't be used to search for extended
emission.  The results of the imaging polarimetry were reported in
\citet{gallagher05}, but the far UV image, which was made with only
F210M filter and no polarimetry elements, was not discussed in that
paper. 

\citet{gallagher05} describe some of the difficulties in identifying
extended emission very close to the point source in an FOC image.
Each filter has its own distinct point-spread function.  For example,
examination of Table 9 in the FOC Instrument
Handbook\footnote{http://www.stsci.edu/hst/foc/documents/handbooks/foc\_handbook.html}
shows that the 210M filter has particularly extended wings compared
with some of the optical filters, with $\sim 20$\% of the flux beyond
$\sim 8$ pixels.  To account for this complication, we
analyzed the FOC PSF image for the F210M 
filter\footnote{http://www.stsci.edu/hst/foc/calibration/f96\_costar.html}
for comparison.  But according to the FOC Instrument Handbook, there
are additional potential problems.  For example, there can be 
geometrical distortion.  Usually well calibrated using the reseau
marks in the larger format (512$\times$512) observations, there may
not be sufficient information for that calibration in the smaller
formats; the Mrk~231 observation was made in the 128$\times$128
format.  Moreover, there is variation in the distortion during the
detector warmup period, and while no data is taken during the initial
warmup period, residual variation on the order of 0.25\% in the plate
scale can be present during the next two hours. According to the
timeline\footnote{http://www.stsci.edu/ftp/observing/weekly\_timeline/1998\_timelines/timeline\_11\_22\_98},
this was the second observation in the sequence.  We can't compare  
with the polarimetry observations, as the polarimetry imaging elements
add their own PSF.   On the other hand, it is not clear that geometric
distortion would be an important effect on the small scales that we
are interested in. Finally, there are also known to be small changes in
telescope focus due to
``breathing''\footnote{http://www.stsci.edu/hst/foc/documents/abstracts/foc\_isr9801.pdf}. 

\begin{figure*}[!t]
\epsscale{1.0}
\begin{center}
\includegraphics[width=3.0truein]{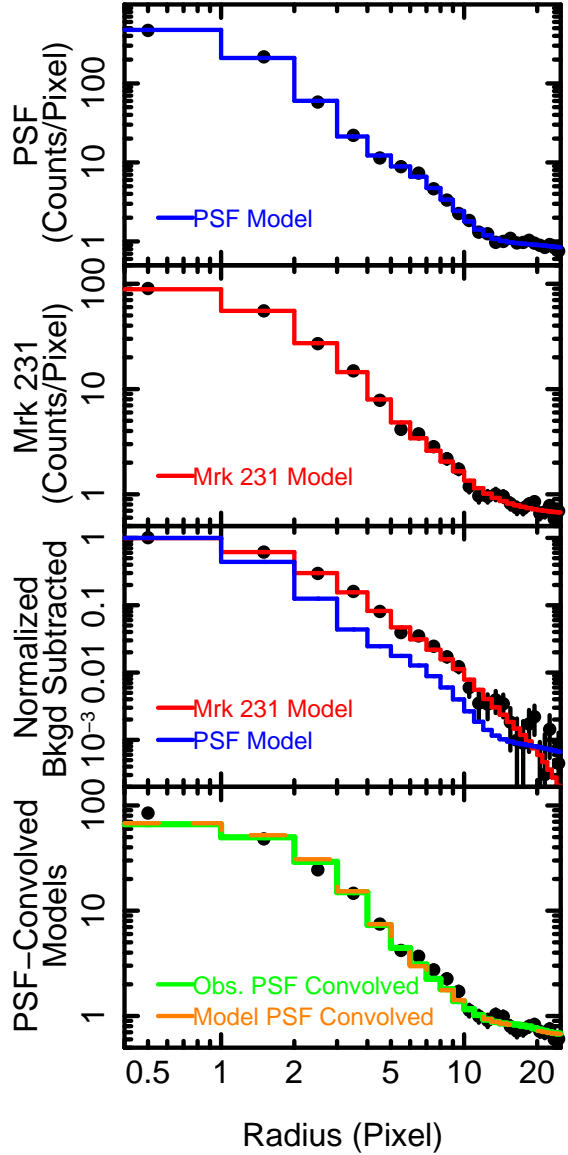}
\caption{Analysis of the {\it HST} FOC F210M observation of Mrk~231.  
 The top panel shows the radial profile of the FOC F210M PSF
 image, fitted with an empirical model consisting of a constant
 background plus four co-aligned 2-D Gaussian models.  The second
 panel shows the results from the same model fit to the  Mrk~231
 image.  The third panel shows  the radial profile for the Mrk~231
 image, constant background  subtracted and normalized to the peak
 value of the PSF model for the  Mrk~231 image.  Overlaid is the PSF model for the PSF observation.
 The radial profile of the PSF is narrower than the Mrk~231 PSF,
 indicating extended emission. 
 The bottom panel shows the radial profile of Mrk~231 and model for a
 fit using a  single Gaussian model convolved with the
 background-subtracted  observed PSF image, and the model PSF
 image. The FWHM of the  Gaussian model is $\sim 3.4$ pixels,
 corresponding to $\sim 40$ pc.   
  Top panel:    \label{fig7}}   
\end{center}
\end{figure*}

We analyzed the Mrk 231 observation and the PSF observation in
parallel using the CXC Ciao {\it Sherpa} package.  We first fit the
two images with co-axial two-dimensional Gaussian profiles plus a
background.  We found that in each case, four Gaussians were
sufficient to empirically describe the PSF.  The radial profiles from
these fits are shown in Fig.~\ref{fig7}.  The rather broad base
corresponds to the particularly extended wings known to characterize
this filter.

The third panel of Fig.~\ref{fig7} shows the empirically-fit radial
profiles for Mrk~231 and the PSF observation, with the constant
background subtracted, normalized, and overlaid on the Mrk~231 radial
profile.  Mrk~231 shows significant excess on the 1--4 pixel scale.  

{\tt Sherpa} offers the capability of fitting the image with a
2-dimensional model after convolving with a PSF image.  We fit the
Mrk~231 observation with a 2-D Gaussian and a constant, using the
background-subtracted PSF observation image as the PSF.  If the
Mrk~231 image were consistent with the detector PSF, the width of the
Gaussian would be equal to zero. Instead, we find a best-fit value of
$3.42 \pm 0.11$ pixels, corresponding to $0.0491 \pm 0.0016$
arcsecond, or 42~pc.  The fit is good, and no distinguishable
improvement in fit is obtained by adding another Gaussian.  The radial
profile of the Gaussian model convolved with the PSF is shown in the
bottom panel of Fig.~\ref{fig7}. 

The FOC PSF is known to not be azimuthally symmetric; moreover the PSF
observation places the PSF star near a reseau mark.  So we also fit the
Mrk~231 image using a PSF developed from the Gaussian-fit model of the
PSF image.  The results were essentially the same; the width of a
single 2-D Gaussian was $3.35 \pm 0.12$ pixels, corresponding to
$0.0481 \pm 0.0017$ arcsecond, or 41~pc. 

Beyond the extended emission, the FOC image provided no exceptional
indication of structure.  For example, a 2-D elliptical model does not
provide notable improvement in fit.  But the signal-to-noise ratio of
this image is low.  The peak pixel contained 93 photons, and we
estimate only $\sim 1600$ source photons for the image within a radius
of 5 pixels, for an average of 20 photons per pixel.  The STIS spatial
analysis suggested that if the offset component were modeled as a
point source, its intensity would be 20\% of the main component.
Simulated images show that such an offset point source might be just
detectable in an image with these statistics.  But the emission might
be truly extended, in which case an asymmetry, lying in the wings of
the nuclear PSF, would be very difficult to detect in an image with
these statistics. 

\section{Discussion\label{discussion}}

\subsection{Summary}

We used observed emission lines from Mrk~231 and a comparison sample to
investigate the binary-black hole model proposed by \citet{yan15}, and
to compare it with the circumstellar absorption model proposed by
\citet{leighly14}.  We used infrared lines, which are subject to
minimal reddening regardless of the model.  The \ion{He}{1}*$\lambda
10830$ line yielded information about the ionization parameter, while
the P$\beta$ line at 12818\AA\/ and the P$\alpha$ line at 18751\AA\/ 
yielded information about the density (due to thermalization at high
density) and column density of the emitting gas.   \ion{C}{4}$\lambda
\lambda 1548, 1551$ was used to probe the ultraviolet, to see whether
that emission line is intrinsically weak and minimally absorbed, as
proposed by \citet{yan15}, or dramatically absorbed, as proposed by
\citet{leighly15}. 

The \citet{yan15} model assumed that the broad line region emission in
Mrk~231 is powered by the photoionizing continuum produced by
thin-disk accretion onto the smaller of the two black holes.   Thus,
in order to produce emission lines in the near infrared that are
seen to have typical equivalent widths with respect to the observed
continuum the equivalent widths with respect small-black-hole-mass
continuum would have to be huge, approximately 100 times larger than
normal, since the photoionizing continuum extrapolated into the
infrared is $\sim100$ times weaker than the observed continuum
(\S\ref{mrk231}).  This seems quite implausible, even without
quantitative analysis, and thus provides the first piece of evidence
that the \citet{yan15} model is untenable.  

Using the photonionization code {\it Cloudy}, we first established
that we were able to produce the observed \ion{He}{1}* equivalent
widths, and \ion{He}{1}*/P$\beta$, P$\alpha$/P$\beta$, and
\ion{He}{1}*/\ion{C}{4} ratios of a comparison sample of objects using
two spectral energy distributions that roughly bracket the properties
of observed AGN and quasar continua.  We  next investigated spectral
energy distributions similar to those proposed by \citet{yan15}, i.e.,
sum-of-blackbodies accretion disks with small inner radii
($r_{in}=3.5r_g$ and Schwartzschild) that were normalized to the
observed far-UV continuum.  We required the X-ray portion of the
spectrum to go through the observed spectrum presented by
\citet{teng14}, a departure from \citet{yan15}, who neglected the
X-ray emission.   Mrk~231 has a rather typical \ion{He}{1}*/P$\beta$
ratio, so we sought simulations that produced both a typical range
of \ion{He}{1}*/P$\beta$ ratios and the observed intensity of the
\ion{He}{1}* emission line. There were none, even if the broad line
region is assumed (unrealistically) to fully cover the continuum
emitting source.  This provided the second piece of evidence that
\citet{yan15} model is untenable.      

Our {\it Cloudy} models showed  that the \ion{He}{1}*/\ion{C}{4} ratio
is sensitive to the SED shape, being lower for harder (X-ray bright)
SEDs and higher for softer (X-ray weak) SEDs, with the maximum values
produced by the SED from the intrinsically X-ray weak quasar PHL~1811.
However, Mrk~231's \ion{He}{1}*/\ion{C}{4} ratio was $\sim 100$ times
higher than  the one from PHL~1811.  We conclude that we do not
see the direct \ion{C}{4}  emission in Mrk~231.  The origin of the
emission that we do observe is not known; we speculate it may be due
to scattering (\S\ref{spec}), or it may be produced in the BAL outflow
\citep{veilleux13, veilleux16}. 

Analysis of the 2-D image from archival {\it HST} STIS observation
shows that the spatial profile of the far-UV continuum is
asymmetrically extended.  Moreover, the spatial profile in the
\ion{Mg}{2} and low-excitation \ion{Fe}{2} troughs is similarly
extended, suggesting that these absorption lines are saturated and the
troughs are filled in by the far-UV continuum.  Evidence for extended
emission in the far-UV was also found in an archival {\it HST} FOC
image.  The scale of the offset emission in both cases was $\sim 40
\rm \, pc$.  

\subsection{Suggestive Results Regarding the Near Infrared Emission
  Lines} 

The focus of this paper is a critique of the \citet{yan15} model, and
we do not purport to provide a full model of the broad line region.
Nevertheless, we have learned a few interesting things about the
infrared lines and {\it Cloudy} models for them.  The observed 
P$\alpha$/P$\beta$ line ratios in the comparison sample are rather low,
with a range between 0.94 and 1.74 and a mean of 1.2.  They are much
lower than the Case A (Case B) values of 2.33 (2.28) in the low density
limit \citep{of06}.  This suggests that the hydrogen-line emission
regions are very optically thick and thermalization is important.
This is not unexpected, as it has been known for many years that the 
Ly$\alpha$/H$\beta$ is observed to be typically around 10, rather than
23 or 34 \citep[the low and high density Case B limits;][]{of06}.
It was surprising, however, to see how thermalization influences line
ratios at different densities to produce a strongly localized density
range for the simple one-zone model.  While arguably not directly
applicable to AGN spectra, it may be useful to consider this
behavior for complete models of the broad line region.  

Also intriguing is the sensitivity of the \ion{He}{1}*/P$\beta$ ratio
to the SED.  On the face of it, this result
is not surprising, given that the \ion{He}{1}* line responds to the
strength of the helium continuum, while the P$\beta$ depends the
ionizing flux and column density of the gas.  This suggests that the
near-infrared broad line ratios could be used to infer the intrinsic
spectral energy distribution of obscured quasars.   But this result
must be viewed with extreme caution.  In this paper, we used a
one-zone model for simplicity, and it is not clear that the same
effect would be present in a extended-BLR model.  Indeed, it is not
clear that these  lines can be trivially modeled using an extended-BLR
model. \citet{ruff12} investigated an LOC model for the hydrogen
lines, and found themselves forced into a small region of photoionization
parameter space. This could be because the standard LOC, which was
historically proposed to explain the high-ionization UV lines 
\citep{baldwin95}, is simply more appropriate for that regime, and
less appropriate for the low-ionization lines considered here.  In
fact, the LOC doesn't always work well for high-ionization lines
\citep{dhanda07}.    In addition, the standard LOC model has a
substantial fraction of optically thin clouds, while we have shown
that high optical depths are needed to produce sufficient line
emission and correct line ratios.  Moreover, it is not clear that the
SED that illuminates low-ionization-line emitting gas is the same
as the continuum we see; it may have been ``filtered'' by gas producing the
high-ionization lines \citep{leighly04}.  Alternative models using
radiation pressure confinement may work better for these lines
\citep{baskin14}.  While this is an interesting and potentially
important problem given the dearth of SED diagnostics the infrared, it
is beyond the scope of the present paper.  

\subsection{Powering the Mid-Infrared Continuum}

While we have demonstrated that the strength of the near-IR broad-line
emission cannot be accounted for by the UV-to-optical SED as observed,
implying the presence of an intrinsically much more luminous ionizing
continuum, there is an additional argument to be made based on
considerations of the mid-infrared power of Mrk~231 that is
independent of the physics of the broad-line region. The thermal near-
and mid-infrared ($2$--$20$~$\mu$m) emission of quasars is attributed
to dust heated by absorption of optical-through-X-ray emission from
the central engine.  The tight, linear correlation between unobscured
optical (0.1--1$\mu$m) and infrared (1--100 $\mu$m) quasar
luminosities seen in radio-quiet quasars supports this interpretation 
\citep[e.g.,][]{gallagher07}. The infrared luminosity is therefore
arguably a more robust indicator of bolometric luminosity than the
optical-UV as observed because the infrared is much less susceptible
to dust extinction.  \citet{gallagher07} recommended the 3$\mu$m
luminosity in particular as a good single value for estimating
quasar bolometric luminosities because the hottest dust is certainly
powered by the AGN with no starburst contamination. The weak PAH
emission seen in the $L$-band and mid-infrared spectra of Mrk~231
supports the claim that this region of the spectrum is dominated by
the quasar \citep{imanishi07,weedman05}.  From the $L$-band spectrum
of Mrk~231 presented in \citet{imanishi07}, the $L_{\rm 3\mu m}$ is
$7.5\times10^{44}$~erg~s$^{-1}$.  Using the 3\micron-to-IR bolometric
correction of $3.44\pm1.68$ \citep{gallagher07} gives $L_{\rm
IR}=(3.86\pm1.26)\times10^{45}$~erg~s$^{-1}$. This is more than an
order of magnitude {\em greater than} the 0.1-1$\mu$m luminosity of
the observed continuum (gray dashed + green dashed continua) shown in
Figure~2: $L_{\rm opt,obs}=2.1\times10^{44}$~erg~s$^{-1}$.  Correcting
for SMC extinction with $E(B-V)=0.1$ following \citet{yan15} brings
$L_{\rm opt,obs}$ up to $2.9\times10^{44}$~erg~s$^{-1}$, still well
below the $L_{\rm IR}$ that is supposedly powered by this continuum.
However, the extinction-corrected 0.1--1.0$\mu$m continuum (orange
dashed line in Fig.~2) gives $L_{\rm
  opt}=2.7\times10^{45}$~erg~s$^{-1}$, in line with the expectations
from the IR power (from Figure~2 of \citealt{gallagher07}). 

\subsection{Polarization}

The \citet{yan15} model also does not
adequately explain the near-UV-to-optical polarization in Mrk~231,
which is significant and strongly rising to the blue through $\sim
3000$\AA\/ \citep{smith95}. 
This kind of polarization signature is commonly seen in reddened
objects.  \citet{wills92} analyzed the similarly-polarized
infrared-luminous quasar IRAS~13349+2438.  They 
showed that electron scattering, which produces a wavelength-independent 
polarization, combined with a reddened continuum produces a
blue-polarized spectrum.  That is, the intrinsic polarization is
constant, but appears to increase toward the blue due to 
dilution by the unpolarized, reddened, direct continuum in the red.
Alternatively, scattering by small dust grains
produces polarization increasing toward the blue (Rayleigh
scattering).  A similar polarization signature is seen among many Type
1 reddened objects  \citep[e.g.,][]{smith00, schmid01, hines01,
  smith02, smith02a,   smith03}.  Mrk~231 is a low-ionization broad
absorption line quasar, and these objects are known to be
significantly polarized  \citep[e.g.,][]{hw95, ogle99, sh99,
  brotherton01, dipompeo11} and  reddened  \citep[e.g.,][]{reichard03,
  dai08, krawczyk15}.   In  addition, Mrk~231 shows evidence for X-ray
absorption \citep{teng14},  and  polarized Type 1 objects are more
likely to suffer X-ray  absorption than unpolarized ones
\citep{leighly97}.    

\subsection{Strategic Absorption\label{strategic}}

\citet{yan15} noted that Mrk~231 has strong optical \ion{Fe}{2}
emission, and therefore might be expected to have comparably strong UV
\ion{Fe}{2} emission, which is not seen.  The  \citet{leighly14}
circumstellar absorption model explains this lack of strong UV
\ion{Fe}{2} naturally via reddening; both the continuum and the line
emission are attenuated in the near UV (Fig.~\ref{fig2}).  Since
\citet{yan15} inferred that reddening is minimal, they proposed 
another explanation for the weak UV \ion{Fe}{2} emission, that since
Mrk~231 is a known \ion{Fe}{2} absorption line quasar
\citep[FeLoBAL,][]{smith95}, the \ion{Fe}{2} emission is absorbed
exactly by the \ion{Fe}{2} absorption in the BAL outflow.  This idea
is untenable for several reasons. First, the velocity offset of the
low-ionization line absorption in Mrk~231 is known to be between about
$-5,500$ and $-4,000 \rm \, km\, s^{-1}$ (e.g., Figure 8 in
\citet{leighly14}). \citet{yan15} required a much broader velocity
width to produce their exact subtraction, between $-8,000$ and
$-1000\rm \, km\, s^{-1}$.   

More importantly, however, the recent {\it HST} STIS observation
(which appears very similar to the spectrum shown in Fig.~\ref{fig2},
albeit with the significant advantage of better signal-to-noise ratio
and  resolution) shows that Mrk 231's near-UV \ion{Fe}{2} absorption
is strong, with saturated low-excitation \ion{Fe}{2} from levels
between 0 and 0.12 eV (near 2600 and 2400 ̊A), while absorption from
higher excitation levels between 0.98 and 1.1 eV (near 2750\AA\/) is
present but weaker {  \citep{veilleux16}.  Some of the
  \citet{leighly14} solutions are therefore ruled out based on the
  presence of   the higher excitation \ion{Fe}{2}, which requires a
  higher density \citep[e.g.,][Fig.\ 13]{lucy14}.  But some of the
  optimized models 
discussed in \S 8 of \citet{leighly14} produce sufficient higher
excitation \ion{Fe}{2}, in particular, the density step-function
models with an illuminated-face density $\log n = 5.5\rm \,
[cm^{-3}]$, constant pressure in the \ion{H}{2} region, and increasing 
by a factor of 25 in the partially ionized zone.  As seen in
Fig.\ 12 in \citet{leighly14}, the inferred location of these
optimized models is 40 parsecs, somewhat interior to the nuclear
starburst, but interestingly consistent with the extended emission
discussed in \S~\ref{fuv}.  Whether or not these models fit the new
STIS data in detail is beyond the scope of the  current paper.}  

\subsection{The Origin of the Resolved FUV Emission\label{spec}}

In \S\ref{fuv}, we describe analysis of archival STIS and FOC data
that indicates the presence of extended far UV continuum emission.
The most convincing evidence comes from the discovery that the spatial
profile is broader and offset in the broad absorption troughs (e.g.,
\ion{Mg}{2}) compared with the near-UV continuum.  The FOC image
suffers from poor signal-to-noise ratio and a complicated PSF, but is
intriguing as the size scale derived ($\sim 40$ parsecs) is basically
the same as that derived from the STIS spatial profile analysis
$>36.5$ parsecs, and a revised estimate of the location of the
absorber.   

One of the mysterious properties of the far UV spectrum is the lack of
broad absorption lines, despite the presence of strong optical,
near-UV, and infrared absorption lines.  This can not be a consequence
of a special combination of photoionization gas parameters.  The
presence of \ion{Fe}{2} absorption means that the column density of
the absorber extends beyond the hydrogen ionization front
\citep[e.g.,][]{lucy14}. The presence of strong \ion{He}{1}*
absorption sets a lower limit on the ionization parameter
\citep{leighly11}.  Strong absorption  from many species would be
expected, including, for example, \ion{C}{4}.  

\citet{veilleux13, veilleux16} explain the lack of far UV absorption
lines via a special geometry
\citep[specifically,][Fig.\ 8]{veilleux16}.  They postulate a dusty 
outflow that covers the optical and near UV emission region, producing
absorption lines on that continuum.  It is opaque to the far UV, but
10\% reaches the viewer unobscured.  The important point, for this
discussion, is that the FUV continuum is thought to be very compact,
emitted by the funnel of a slim disk, and it would be expected to be
unresolved.  

The presence of extended emission casts some doubt upon this
explanation. Our analysis shows that a significant amount of FUV
continuum is not coincident with the central engine emitting the
near-UV and optical continuum.  It is possible that all of the far-UV
continuum emerging from the central engine is attenuated by the
significant reddening that is the origin of the rolloff in the optical
and UV.  Then, the observed far-UV continuum comes from somewhere near
the central engine, but not from the central engine, and that is why
there are no far-UV broad absorption lines.

{ UV variability is commonly observed in AGN and quasars, and it
  can be used to identify them  \citep[e.g.,][]{peters15,
    morganson16}.  An origin of the far UV continuum in extended 
  emission in Mrk~231 predicts  that the continuum should not be 
  variable.    \citet{veilleux16}  report no significant FUV
  variability  between two COS observations   separated by 3 years.
  The lack of variability provides further  support for    dominance
  of an   extended emission component for the far UV   continuum.  }

The offset emission may be somewhat similar to an optical continuum
peak or ``hot spot'' in the inner narrow-line region of the nearby
Seyfert 2 galaxy NGC~1068.  Long-slit STIS spectroscopy revealed a
continuum shape indistinguishable from Seyfert 1 galaxies, as well as
broad components of emission lines such as \ion{C}{4}.  \citet{ck00} 
conclude that this component is reflected emission from the central
engine.  The hot spot lies 30 pc from the estimated position of the
central engine in NGC~1068 \citep{kc_00}, intriguingly close to the
estimates of the location of the extended emission in Mrk~231.
An important difference, from a data analysis point of view, is that
NGC~1068 is much nearer than Mrk~231; e.g., an arcsecond corresponds
to 60 parsecs in NGC~1068 versus 867 parsecs in Mrk 231.

We suggest that a scattering /reflection scenario may have been too
hastily dismissed by \citet{veilleux13}.  The continuum flux near
2000\AA\/ is found to be about 165 times weaker than the intrinsic
continuum inferred by \citet{leighly14}, i.e., 0.6\% of the intrinsic
flux.  Such a level of scattering is not implausible; e.g., a median 
scattering efficiency of 2.3\% is found in a sample of obscured
quasars \citep[e.g.,][although the scattering regions in those 
galaxies are very large]{obied15}. \citet{veilleux13} dismiss
scattering due to the low level of polarization in the UV, as observed
by \citet{smith95}, but we note that while the polarization 
in the UV is lower than it is in the near-UV, it is not zero.
Moreover, it shows a position angle rotation compared with the strong
near-UV polarization, which suggests a different
origin. Also, we note that the degree of polarization that is observed
depends on the asymmetry of the scatterer.  For example, the hot spot
in NGC~1068 shows a high degree of polarization, but the geometry in
that case (a relatively small, single reflector well resolved from the
hidden nucleus) is nearly ideal for producing high polarization.  If
the scatterer in Mrk~231 has a large solid angle to the nucleus, high
polarization is not expected.  In addition, the optical polarization 
in Mrk~231 has been observed to be variable \citep{gallagher05}, and
the {\it HST} UV polarization observation was performed more than 20
years ago.   

Finally, an intriguing possibility is that the far UV continuum may be 
nuclear continuum reflected from the wind on the far side of the
nucleus.  This is suggested by the confluence of distance estimates:
36-40 parsecs for the extended emission (\S~\ref{fuv}), and $\sim 40$
parsecs as a revised approximate distance to the broad absorption line
gas (\S~\ref{strategic}).  

\acknowledgements

KML acknowledges useful conversations with Dick Henry, { Mike
Brotherton, and Jerry Kriss.  She also thanks Mike Crenshaw for
reviewing the imaging/spatial analysis section, and suggesting the 
intriguing parallel with NGC 1068, and the anonymous referee for
suggestions that improved the content and clarity of the paper.  ABL
thanks Zoltan Haiman for useful comments.   The authors gratefully
acknowledge Jules Halpern for the title suggestion.}  KML thanks 
Erin Cooper for processing the Mrk 231 spectra.  KML acknowledges
support through STScI HST-GO-14058.001-A. SCG thanks the Natural
Science and Engineering Research Council of Canada for support.

{\it Facilities:}  \facility{IRTF (SpeX)}, \facility{HST (STIS, FOC)}

\end{document}